\documentclass[a4paper,11pt,notoc]{article}
\usepackage{jheppub}
\usepackage{float}
\usepackage{subfig}
\usepackage{ulem}
\usepackage[table]{xcolor}
\usepackage[compat=1.1.0]{tikz-feynman}
\newcommand{\be}{\begin{equation}}
\newcommand{\ee}{\end{equation}}
\def\bsp#1\esp{\begin{split}#1\end{split}}
\def\bpm{\begin{pmatrix}}
\def\epm{\end{pmatrix}}

\preprint{LAPTH-024/21, TTP21-022, P3H-21-050}

\title{Dark matter and lepton flavour phenomenology in a singlet-doublet scotogenic model}

\author[a]{Maud Sarazin}
\author[b,c]{~Jordan Bernigaud}
\author[a]{~Bj\"orn Herrmann}

\affiliation[a]{Univ.\ Grenoble Alpes, Univ.\ Savoie Mont Blanc, CNRS, LAPTh, F-74000 Annecy, France}
\affiliation[b]{Institute for Nuclear Physics (IKP), Karlsruhe Institute of Technology, Hermann-von-Helmholtz-Platz 1, D-76344 Eggenstein-Leopoldshafen, Germany}
\affiliation[c]{Institute for Theoretical Particle Physics (TTP), Karlsruhe Institute of Technology, Engesserstrasse 7, D-76128 Karlsruhe, Germany}

\emailAdd{sarazin@lapth.cnrs.fr}
\emailAdd{jordan.bernigaud@kit.edu}
\emailAdd{herrmann@lapth.cnrs.fr}

\abstract{We study the dark matter phenomenology of scotogenic frameworks through a rather illustrative model extending the Standard Model by scalar and fermionic singlets and doublets. Such a setup is phenomenologically attractive since it provides the radiative generation of neutrino masses, while also including viable candidates for cold dark matter. We employ a Markov Chain Monte Carlo algorithm to explore the associated parameter space in view of numerous constraints stemming from the Higgs mass, the neutrino sector, dark matter, and lepton-flavour violating processes. After a general discussion of the results, we focus on the case of fermionic dark matter, which remains rather uncovered in the literature so far. We discuss the associated phenomenology and show that in this particular case a rather specific mass spectrum is expected with fermion masses just above 1 TeV. Our study may serve as a guideline for future collider studies.}


\begin{document}
\notoc
\maketitle
\flushbottom

% ===========================================================================
\section{Introduction}
\label{Sec:Introduction}
% ======================================================

Particle physics beyond the Standard Model (BSM) is well motivated by many arguments from both theory considerations and experimental observations. 

Among the latter, three important observations clearly call for new physics. First, the existence of cold dark matter (DM) in the Universe \cite{Planck} remains unexplained. Second, the discovery of their oscillations \cite{KamLand} implies that the neutrinos are massive, although there is no corresponding mass term in the Standard Model. Finally, the baryon asymmetry observed in the Universe \cite{Planck} cannot be explained without invoking physics beyond the Standard Model.

One of the classical approaches to address dark matter relies on the so-called Weakly Interacting Massive Particle (WIMP) paradigm. Such a massive, electrically neutral, and stable particle may achieve the observed dark matter relic density through thermal freeze-out. However, in absence of direct experimental evidence, there is no consensus about the exact nature of the WIMP dark matter candidate.

Concerning neutrino masses, while in the Standard Model the other fermions are of Dirac nature, allowing for mass generation through the Higgs mechanism, right-handed neutrinos would be a singlet under full Standard Model gauge group. In addition, accounting for the experimentally inferred mass scale would require unnaturally small Yukawa couplings. While many extensions of the Standard Model address this question by considering effective small masses generated through the so-called Seesaw mechanism \cite{Minkowski:1977sc, Yanagida:1980xy, PhysRevLett.56.561}, a different approach can be implemented by considering loop-induced Majorana mass terms, which are naturally loop-suppressed.

In the present work, we will consider a framework of the scotogenic type, which reunites WIMP dark matter and radiative neutrino mass generation. A simple model of this type has been proposed by Ma \cite{Ma:2006km} and discussed in Refs.\ \cite{Toma:2013zsa, Vicente:2014wga, Fraser:2014yha, Baumholzer:2019twf}. Phenomenological aspects of this class of models have then been presented in numerous subsequent studies \cite{Klasen:2013jpa, Molinaro:2014lfa, Lindner:2016kqk, Rocha-Moran:2016enp, Ahriche:2017iar, Bhattacharya:2017sml, Bhattacharya:2018fus, Ahriche:2018ger, Konar:2020wvl, Escribano:2020iqq, DeRomeri:2021yjo}. A summary and classification in terms of topologies related to neutrino mass generation of more general scotogenic setups has been published in Ref.\ \cite{Restrepo2013}.

The generation of neutrino masses necessarily comes along with lepton-flavour violating effects, which may contribute to experimentally constrained processes, such as, e.g., the decay $\mu \to e\gamma$ and related leptonic transitions. Precision measurements of such transitions may put stringent constraints on the parameter space, especially concerning coupling parameters involved in the neutrino mass generation and dark matter phenomenology. Moreover, the same couplings may also contribute to a processes involved in leptogenesis \cite{Fukugita:1986hr} allowing to explain the baryon asymmetry through an initial lepton asymmetry. While, in the present paper, we will address dark matter and neutrino masses, this last point is left for future work.

We will dedicate our analysis to a specific scotogenic model, labelled ``T1-2A'' according to Ref.\ \cite{Restrepo2013} that includes both scalar and fermionic singlets and doublets. This model features two non-vanishing neutrino masses generated through radiative contributions and offers three different candidates for WIMP dark matter. A discrete $\mathbb{Z}_2$ symmetry ensures that the dark matter particle is stable. In a sense, this corresponds to a union of the singlet-doublet scalar \cite{Cohen2011, Cheung2013, Banik2014, Cabral2017} and singlet-doublet fermion \cite{Cohen2011, Cheung2013, Calibbi2015, Bhattacharya:2015qpa, Banerjee2016, Abe2017} extensions. More recently, this model has been studied in Ref.\ \cite{Esch2018}, focusing mainly on the particular case of scalar dark matter.

The present work aims at providing a more complete analysis of this rather general framework. In particular, we do not restrict ourselves to the case of scalar dark matter. Moreover, we include $CP$-violating phases present in the Pontecorvo-Maki-Nakagawa-Sakata (PMNS) matrix. Finally, we include the couplings of the new fields to the right-handed leptons, which have been omitted in Ref.\ \cite{Esch2018} as they do not contribute to the generation of neutrino masses. They may, however, have an impact on lepton flavour violating processes and dark matter phenomenology. In order to efficiently include all relevant constraints and explore the large parameter space, we rely on a Markov Chain Monte Carlo algorithm. To our knowledge, this is the first application of this technique to models aiming at explaining neutrino mass generation.

Our paper is organised as follows: After a presentation of the model in Sec.\ \ref{Sec:Model}, we discuss the physical mass spectrum as well as the impact of next-to-leading order corrections on the masses in Sec.\ \ref{Sec:Spectrum}. Sec.\ \ref{Sec:Setup} is then devoted to the presentation of the employed Markov Chain Monte Carlo setup. We present our results in Sec.\ \ref{Sec:Results} and discuss perspectives for collider searches in Sec.\ \ref{Sec:Outlook}. Finally, conclusions are given in Sec.\ \ref{Sec:Conclusion}.
% ======================================================
\section{The model T1-2A}
\label{Sec:Model}
% ======================================================

As mentioned in the Introduction, we will consider a scotogenic framework, where the Standard Model is extended by two Weyl fermion doublets, $\Psi_1$ and $\Psi_2$, a Weyl fermion singlet, $F$, a scalar doublet, $\Phi$, and a real scalar singlet, $S$. These additional fields are assumed to be singlet under $SU(3)$, and odd under a $\mathbb{Z}_2$ symmetry to ensure neutrino mass generation at the one loop level as well as the stability of the dark matter candidate. All Standard Model fields are even under the mirror symmetry. The additional field content including their respective representations under $SU(2)_L\times U(1)_Y$ is summarized in Table \ref{Tab:T12AQuantumNumbers}. According to the classification of Ref.\ \cite{Restrepo2013}, this specific scotogenic model is labelled ``T1-2A'', where the ``T1-2'' refers to the topology of adding two fermions and two scalars, while the ``A'' indicates that the model contains two singlets and two doublets.

\begin{table}
    \centering
    \begin{tabular}{|c||c|c|c||c|c|}
    \hline
            \     & ~$\Psi_{1}$~ & ~$\Psi_2$~ & ~$F$~ & ~$\Phi$~ & ~$S$~ \\
            \hline
            \hline
      $SU(2)_L$   &  $\mathbf{2}$ & $\mathbf{2}$ & $\mathbf{1}$ & $\mathbf{2}$ & $\mathbf{1}$ \\
      \hline
        $U(1)_Y$  & -1 & 1 & 0 & 1 & 0 \\
        \hline
    \end{tabular}
    \caption{Field content of the scotogenic model T1-2A in addition to the Standard Model fields.}
    \label{Tab:T12AQuantumNumbers}
\end{table}

In this Section, we will briefly introduce the different sectors, present the corresponding Lagrangian, and set the notation. 

% ======================================================
\subsection{The scalar sector}
\label{Sec:ScalarSector}
% ======================================================

The scalar sector of the model consists of the Standard Model Higgs doublet $H$, the additional singlet $S$, and the additional doublet $\Phi$. They carry charges as given in Table \ref{Tab:T12AQuantumNumbers}. Upon electroweak symmetry breaking, which involves the Higgs doublet only, the doublets can be expanded into components according to
\begin{align}
	H ~=~ \begin{pmatrix} G^+ \\ \frac{1}{\sqrt{2}} \big[ v + h^0 + i G^0 \big] \end{pmatrix}, \qquad
	\Phi ~=~ \begin{pmatrix} \phi^+ \\ \frac{1}{\sqrt{2}} \big[ \phi^0 + i A^0 \big] \end{pmatrix} \,.
\end{align}
Here, $h^0$ is the Standard-Model Higgs boson, $G^0$ and $G^+$ are the Goldstone bosons, and $v = \sqrt{2} \langle H \rangle \approx 246$ GeV denotes the vacuum expectation value. Moreover, $\phi^0$ and $A^0$ are $CP$-even and $CP$-odd neutral scalars, and $\phi^+$ is a charged scalar. 

The scalar potential of the model is given by
\begin{align}\begin{split}
	-{\cal L}_{\rm scalar} ~=&~ M_H^2 \big| H \big|^2 + \lambda_H \big| H \big|^4 + \frac{1}{2} M_S^2 S^2 + \frac{1}{2} \lambda_{4S} S^4 + M_{\Phi}^2 \big| \Phi \big|^2 + \lambda_{4\Phi} \big| \Phi \big|^4 + \frac{1}{2} \lambda_S S^2 \big| H \big|^2 \\ &~+ \lambda_{\Phi} \big| \Phi \big|^2 \big| H \big|^2 
		+ \lambda'_{\Phi} \big| H \Phi^{\dag} \big|^2 + \frac{1}{2} \lambda''_{\Phi} \Big\{ \big( H \Phi^{\dag} \big)^2 + {\rm h.c.}\Big\} + T \Big\{ S H \Phi^{\dag} + {\rm h.c.} \Big\} \,.
	\label{Eq:ScalarPotential}
\end{split}\end{align}
The first two terms are the Standard Model part related to the Higgs doublet $H$. After electroweak symmetry breaking, at the tree-level, the usual minimization relation in the Higgs sector,
\begin{align}
	m^2_{h^0} ~=~ -2 M_H^2 ~=~ 2 \lambda_H v^2 \,,
	\label{Eq:HiggsMass}
\end{align}
allows to eliminate the mass parameter $M_H^2$ in favour of the Higgs self-coupling $\lambda_H$. Imposing $m_{h^0} \approx 125$ GeV leads to a tree-level value of $\lambda_H \approx 0.13$. 

The additional couplings, stemming from the presence of the new fields, are the self-couplings $\lambda_{4S}$ and $\lambda_{4\Phi}$ of the new singlet $S$ and doublet $\Phi$, the coupling $\lambda_S$ between the Standard Model Higgs doublet $H$ and the new singlet $S$, the couplings $\lambda_{\Phi}$, $\lambda_{\Phi'}$, and $\lambda_{\Phi''}$ between the Standard Model Higgs doublet $H$ and the new doublet $\Phi$, and the trilinear coupling $T$ relating the Higgs doublet $H$, the new singlet $S$, and the new doublet $\Phi$. Moreover, there are the mass terms $M_S^2$ and $M_{\Phi}^2$ for the singlet and doublet, respectively. 

The resulting physical scalar mass eigenstates will be discussed in Sec.\ \ref{Sec:ScalarMasses}.

% ======================================================
\subsection{The fermion sector}
\label{Sec:FermionSector}
% ======================================================

In addition to the Standard Model fermions, the model contains a Majorana singlet $F$ and a Dirac doublet $\Psi$. The latter can be expressed as two Weyl components with opposite hypercharge,
\begin{align}
	\Psi_1 ~=~ \begin{pmatrix} \Psi^0_1 \\ \Psi^-_1 \end{pmatrix} \,, \qquad
	\Psi_2 ~=~ \begin{pmatrix} -\big( \Psi^-_2 \big)^{\dag} \\ \big( \Psi^0_2 \big)^{\dag} \end{pmatrix} \,.
\end{align}

The corresponding Lagrangian includes mass terms $M_F$ and $M_{\Psi}$ for the singlet and the doublets, respectively, and Yukawa couplings $y_1$ and $y_2$ inducing mixing between the singlet and one of the doublets,
\begin{align}\begin{split}
	-{\cal L}_{\rm fermion} ~=~& -\frac{i}{2} \Big( \overline{\Psi}_1 \sigma^{\mu}D_{\mu} \Psi_1 + \overline{\Psi}_2 \sigma^{\mu} D_{\mu} \Psi_2 \Big) \\
	&+ \frac{1}{2} M_F F^2 + M_{\Psi} \Psi_1 \Psi_2 + y_1 \Psi_1 H F + y_2 \overline{\Psi}_2 H \overline{F} + {\rm h.c.}
	\label{Eq:LagFermions}
\end{split}\end{align}
The physical mass eigenstates are three neutral fermions, potentially a superposition of singlet and doublet, as well as one charged fermion. The fermionic mass spectrum will be discussed in more detail in Sec.\ \ref{Sec:FermionMasses}.

% ======================================================
\subsection{Interaction terms and neutrino masses}
\label{Sec:Interaction}
% ======================================================

\begin{figure}
    \centering
    \includegraphics[scale=1.0]{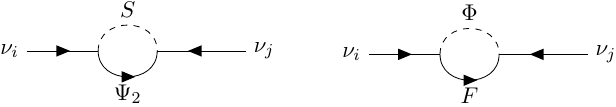}
	\caption{Feynman diagrams associated to the generation of neutrino masses at the one-loop level in the model T1-2A in the interaction eigenbasis.}
\label{Fig:RadiativeNeutrinoMasses}
\end{figure}

The Lagrangian of the model is completed by interaction terms that relate the scalar and fermion sectors to the Standard Model leptons. They are given by
\begin{align}
    -{\cal L}_{\rm interaction} ~=~ g_{\Psi}^i \Psi_2 L_i S + g_F^i \Phi L_i F + g_R^i E^c_i \Phi^{\dag} \Psi_1 \,,
	\label{Eq:LagCouplings}
\end{align}
where $L_i$ ($i=e,\mu,\tau$) denotes the three generations of lepton doublets in the Standard Model, and $E_i$ the singlets of right-handed charged leptons. The presence of the first two terms of Eq.\ \eqref{Eq:LagCouplings} allows the radiative generation of neutrino masses, and constitutes an important feature of the model T1-2A with respect to the extensions of the Standard Model by only scalars \cite{LopezHonorez:2006gr, LopezHonorez:2010eeh, LopezHonorez:2010tb, Cohen2011, Cheung2013, IDM2013, Banik2014, Cabral2017}, only fermions \cite{Cohen2011, Cheung2013, Calibbi2015, Banerjee2016, Abe2017}, or models with only fermionic and scalar singlets \cite{Esch:2014jpa}. The corresponding Feynman diagrams in the interaction basis are shown in Fig.\ \ref{Fig:RadiativeNeutrinoMasses}. 

For practical reasons, we regroup the couplings appearing in the first two terms of Eq.\ \eqref{Eq:LagCouplings} into the matrix
\begin{equation}
    {\cal G} ~=~ 
        \begin{pmatrix} 
            g_{\Psi}^e & g_{\Psi}^{\mu} & g_{\Psi}^{\tau} \\[1mm]
            g_F^e & g_F^{\mu} & g_F^{\tau}
        \end{pmatrix}
    \label{Eq:CouplingMatrix}
\end{equation}
Note that the last term of Eq.\ \eqref{Eq:LagCouplings} involving the couplings $g_R^i$ does not contribute to neutrino masses, but may have implications in other sectors, in particular in lepton-flavour violating and dark-matter related processes.

% ======================================================
\section{Physical mass spectrum}
\label{Sec:Spectrum}
% ======================================================

Having discussed the Lagrangian parameters in the previous Section, we will now focus on the resulting physical mass spectrum. We will in particular discuss the impact of next-to-leading order (NLO) corrections to the physical masses of scalars and fermions. We will also present the different candidates for cold dark matter within the model T1-2A.  

% ======================================================
\subsection{Scalar masses}
\label{Sec:ScalarMasses}
% ======================================================

After electroweak symmetry breaking, the neutral scalar states $S$ and $\phi^0$ mix, giving rise to two $CP$-even neutral mass eigenstates, which we will denote as $\phi^0_1$ and $\phi^0_2$. At the tree-level, in the basis $\{ S, \phi^0 \}$ the corresponding mass matrix is given by 
\begin{align}
	{\cal M}^2_{\phi} ~=~ \begin{pmatrix} M_S^2 + \frac{1}{2}v^2\lambda_S & v T \\ v T & M_{\Phi}^2 + \frac{1}{2} v^2 \lambda_L \end{pmatrix} \,,
\end{align}
where $\lambda_L = \lambda_{\Phi} + \lambda'_{\Phi} + \lambda''_{\Phi}$. Diagonalization according to 
\begin{equation}
    \big( \phi^0_1, \phi^0_2 \big) = U_{\phi} \, \big( S, \phi^0 \big)
\end{equation} 
leads to the mass eigenvalues
\begin{align}
    \begin{split}
	m^2_{\phi^0_{1,2}} ~=~& \frac{1}{2} \Biggr[ M_S^2 + M_{\Phi}^2 + \frac{1}{2}v^2 \big( \lambda_S + \lambda_L \big)
	\mp \sqrt{ \Big[ M_S^2 - M_{\Phi}^2 + \frac{1}{2}v^2 \big( \lambda_S - \lambda_L \big) \Big]^2 + 4 v^2 T^2} \Biggr] \,,
    \end{split}
\end{align}
defined such that $m_{\phi^0_1} < m_{\phi^0_2}$. Finally, the masses of the $CP$-odd and charged scalars are, at the tree-level, given by
\begin{align}
    \begin{split}
	m^2_{A^0} ~=~ M_{\Phi}^2 + \frac{1}{2}v^2 \Big[ \lambda_{\Phi} + \lambda'_{\Phi} - \lambda''_{\Phi} \Big] \qquad {\rm and} \qquad
	m^2_{\phi^\pm} ~=~ M_{\Phi}^2 + \frac{1}{2}v^2 \lambda_{\Phi} \,.
	\end{split}
\end{align}

In our analysis, we will make use of the numerical spectrum calculator {\tt SPheno} (version {\tt 4.0.4}) \cite{SPheno2003, SPheno2012} to compute the mass spectrum including radiative corrections at the one-loop level. To this end, we have implemented the T1-2A model under consideration into the {\tt Mathematica} package {\tt SARAH} \cite{SARAH2010, SARAH2011, SARAH2013, SARAH2014} allowing to generate the numerical modules for {\tt SPheno}. In the following, we will discuss the impact of radiative corrections on the scalar mass spectrum.

\begin{figure}
    \centering
    \includegraphics[width=0.49\textwidth]{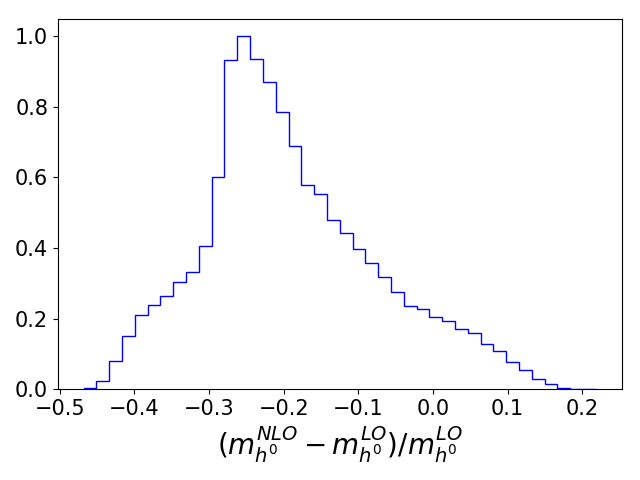}~~~
    \includegraphics[width=0.49\textwidth]{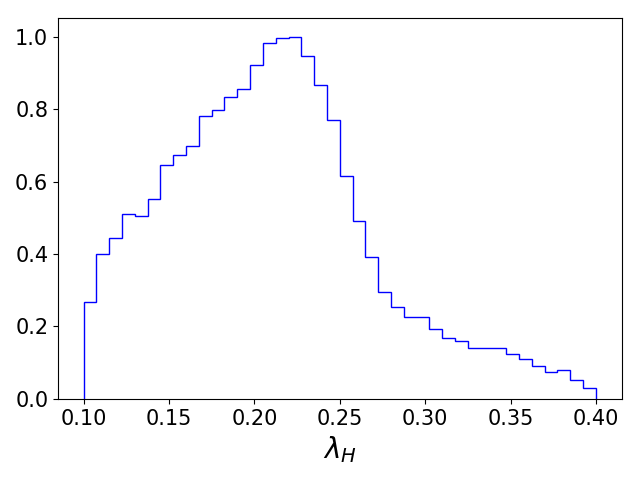}
    \vspace*{-8mm}
    \caption{Left: Histogram showing the relative impact of the higher-order corrections to the Higgs-boson mass. Right: Resulting distribution of the tree-level parameter $\lambda_H$ when imposing the measured Higgs-boson mass. Both distributions are based on a sample of about 80.000 phenomenologically viable parameter points.}
    \label{Fig:HiggsMass}
\end{figure}

The mass of the Standard Model Higgs boson $h^0$ is mainly governed by the parameter $\lambda_H$ (see Eq.\ \eqref{Eq:HiggsMass}), while the remaining parameters of the scalar sector enter only through the one-loop corrections. In the left part of Fig.\ \ref{Fig:HiggsMass} we show the distribution of the relative correction to the tree-level mass, obtained for a sample of about 94.150 phenomenologically viable parameter configurations. Our exact setup will be detailed in Sec.\ \ref{Sec:Setup}. As can be seen, the corrections to the Higgs mass may be sizeable, reaching about 45\% in extreme cases. Consequently, the viable interval for the parameter $\lambda_H$ is widened with respect to the pure tree-level case, where $\lambda_H \approx 0.13$. The resulting distribution for $\lambda_H$ is shown on the right part of Fig.\ \ref{Fig:HiggsMass} justifying the interval $\lambda_H \in [0.1; 0.4]$ for the following study.

\begin{figure}
    \centering
    \includegraphics[width=0.49\textwidth]{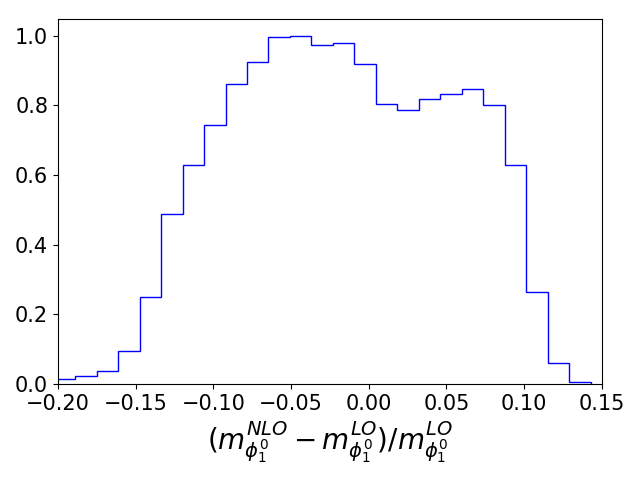}~~~
    \includegraphics[width=0.49\textwidth]{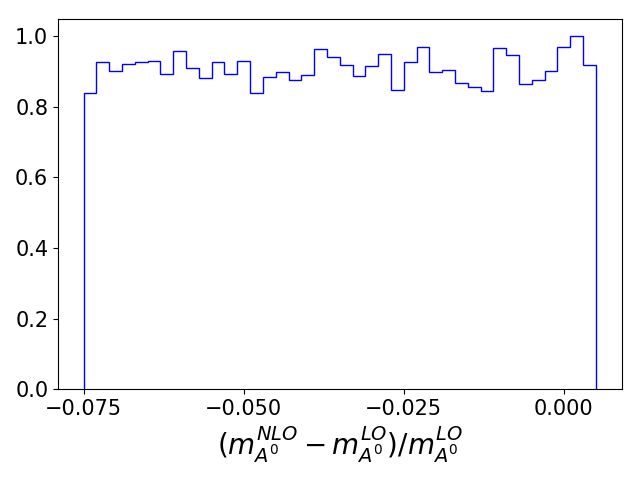}
    \vspace*{-8mm}
    \caption{Histograms showing the relative impact of the higher-order corrections for the lighter scalar ($m_{\phi^0_1}$, left) and the pseudo-scalar ($m_A^0$, right) obtained for the same parameter sample as in Fig.\ \ref{Fig:HiggsMass}.}
    \label{Fig:ScalarMass}
\end{figure}

Coming to the additional scalar states, Fig.\ \ref{Fig:ScalarMass} shows distribution of the correction to the lighter scalar and the pseudo-scalar masses, obtained for the same parameter space sample. Again, different corrections arise from the different values of scalar and fermion parameters. They can either increase or decrease the mass, however to a much lesser extent than for the Higgs-boson discussed above. Here, typical corrections are of the order of a few GeV in most cases, corresponding to a relative correction of up to about 10\%. The corrections are of the same order for the heavier scalar $\phi^0_2$.

Overall, the corrections in this sector are similar to those in two-Higgs doublet models, such as the Inert Doublet Model \cite{LopezHonorez:2006gr, LopezHonorez:2010eeh, LopezHonorez:2010tb, IDM2013}, the additional singlet state does not affect the corrections in a significant way.

% ======================================================
\subsection{Fermion masses}
\label{Sec:FermionMasses}
% ======================================================

In the fermion sector, the singlet $F$ will mix with the neutral doublet components  $\Psi_1$ and $\Psi_2$. At the tree-level, in the basis $\big\{ F, \Psi^0_1, \Psi_2^{0\dagger} \big\}$, the corresponding mass matrix is given by
\begin{align}
	{\cal M}_{\chi^0} ~=~ \begin{pmatrix} 
			M_F & \frac{v}{\sqrt{2}} y_1 & \frac{v}{\sqrt{2}} y_2 \\
			\frac{v}{\sqrt{2}} y_1 & 0 & M_{\Psi} \\
			\frac{v}{\sqrt{2}} y_2 & M_{\Psi} & 0
	 	\end{pmatrix} 
	 	~=~ \begin{pmatrix} 
	 	    M_F & \frac{v y}{\sqrt{2}} \cos\theta & \frac{v y}{\sqrt{2}} \sin\theta \\
	 	    \frac{v y}{\sqrt{2}} \cos\theta & 0 & M_{\Psi} \\
	 	    \frac{v y}{\sqrt{2}} \sin\theta & M_{\Psi} & 0 
	 	\end{pmatrix} \,.
\end{align}

The resulting physical mass eigenstates are denoted $\chi^0_1$, $\chi^0_2$, and $\chi^0_3$. The associated mixing matrix is defined according to
\begin{align}
	\big( \chi^0_1, \chi^0_2, \chi^0_3 \big) ~=~ U_{\chi} \, \big( F, \Psi^0_1, \Psi^0_2 \big) \,,
\end{align}
ordered by definition such that $m_{\chi^0_1} < m_{\chi^0_2} < m_{\chi^0_3}$.

\begin{figure}
    \centering
    \includegraphics[width=0.49\textwidth]{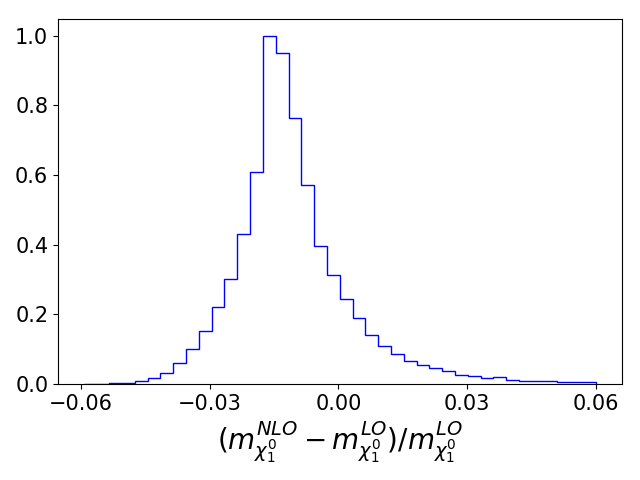}~~~
    \includegraphics[width=0.49\textwidth]{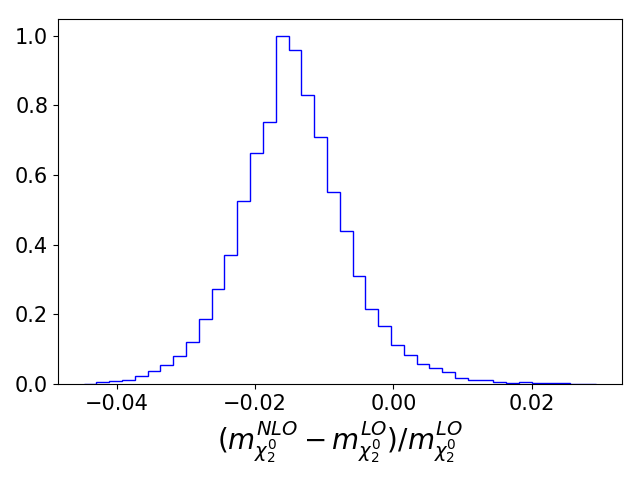}
    \vspace*{-8mm}
    \caption{Histogram showing the relative impact of the higher-order corrections for the lightest ($m_{\chi^0_1}$, left) and the second-lightest ($m_{\chi^0_2}$, right) fermions obtained for the same parameter sample as in Fig.\ \ref{Fig:HiggsMass}.}
    \label{Fig:FermionMass}
\end{figure}

As for the scalar masses, we analyse the impact of the one-loop corrections included in {\tt SPheno} by comparing to the tree-level results for the fermion masses. We show the example of the lightest and the second-lightest fermion state in Fig.\ \ref{Fig:FermionMass}. Typically, the correction to the lighter mass is of the same order as for the new scalars discussed above, the corrections not exceeding about 2\% for a large majority of points. The correction received by the second-lightest state is slightly less important. The impact on the heaviest fermion mass (not shown in Fig.\ \ref{Fig:FermionMass}) is similar to the one on the second state.

Let us note that, although the absolute impact of the one-loop corrections turns out to be numerically small, it may have a signification impact when considering co-annihilations in the context of computing the dark matter relic density. The contributions from co-annihilation are extremely sensitive to mass differences, which may be affected by the presented one-loop corrections.

% ======================================================
\subsection{Neutrino masses}
\label{Sec:NeutrinoMasses}
% ======================================================

\begin{figure}
    \centering
    \includegraphics[scale=1.0]{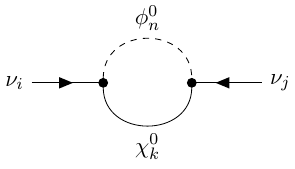}
	\caption{Feynman diagram associated to the generation of neutrino masses at the one-loop level in the mass eigenbasis. Here, the neutrino indices $i, j$ run over $e, \mu, \tau$, the scalar $\phi^0_n$ can be $CP$-even ($\phi^0_1$, $\phi^0_2$) or $CP$-odd ($A^0$), and the fermionic states include $\chi^0_1$, $\chi^0_2$, and $\chi^0_3$.}
	\label{Fig:RadiativeNeutrinoMasses2}
\end{figure}

After electroweak symmetry breaking, the neutrino masses receive contributions from the three neutral Majorana fermions $\chi^0_1$, $\chi^0_2$, and $\chi^0_3$ together with the neutral scalars $\phi^0_1$, $\phi^0_2$, and $A^0$. The neutrino masses are radiatively generated via the diagrams shown in Fig.\ \ref{Fig:RadiativeNeutrinoMasses2} involving the couplings $g_F$ and $g_{\Psi}$. The resulting neutrino mass matrix can be expressed as
\begin{equation}
    {\cal M}_{\nu} ~=~ {\cal G}^t \, M_L \, {\cal G} \,,
    \label{Eq:NeutrinoMassMatrix}
\end{equation}
where the couplings ${\cal G}$ are separated from the loop functions contained in the matrix $M_L$. The latter depends only on the parameters of the scalar and fermion sectors. Due to the precise field content of the present model, there are only two independent contributions to the neutrino mass. Consequently, the matrix $M_L$ is of order $2 \times 2$ meaning that the present framework can only provide two non-zero neutrino masses. Detailed expressions for the elements of $M_L$ are given in App.\ \ref{App:NeutrinoMassMatrix}.

The neutrino mass eigenvalues are obtained by diagonalizing the mass matrix ${\cal M}_{\nu}$ such that the physical mass eigenstates are given by
\begin{align}
    \left( \nu_1, \nu_2, \nu_3 \right) ~=~ 
        U_{\nu} \left( \nu_e, \nu_{\mu}, \nu_{\tau} \right) \,,
    \label{Eq:NeutrinoDiagonalization}
\end{align}
again ordered such that $m_{\nu_1} \equiv 0 < m_{\nu_2} < m_{\nu_3}$. 

In the basis where the charged lepton Yukawa matrix is diagonal, the mixing matrix $U_{\nu}$ in Eq.\ \eqref{Eq:NeutrinoDiagonalization} is identified with the Pontecorvo-Maki-Nakagawa-Sakata (PMNS) matrix \cite{Pontecorvo1957b, Maki1962}. The latter can be expressed as the product
\begin{align}
    U_{\rm PMNS} ~=~ \!\!
		\begin{pmatrix} 
		    1 & 0 & 0 \\ 0 & c_{23} & s_{23} \\ 0 & -s_{23} & c_{23} 
		\end{pmatrix} \!\!
		\begin{pmatrix} 
		    c_{13} & 0 & s_{13} e^{-i \delta_{\rm CP}} \\ 
		    0 & 1 & 0 \\ 
		    -s_{13}e^{i \delta_{\rm CP}} & 0 & c_{13} 
		\end{pmatrix} \!\!
		\begin{pmatrix} 
		    c_{12} & s_{12} & 0 \\ -s_{12} & c_{12} & 0 \\ 0 & 0 & 1 
		\end{pmatrix} \!\!
		\begin{pmatrix} 
		    1 & 0 & 0 \\ 
		    0 & e^{i \alpha_{1}} & 0 \\ 
		    0 & 0 & e^{i \alpha_{2}}
		\end{pmatrix} \,,
	\label{Eq:PMNSMatrix}
\end{align}
where $\theta_{12}$, $\theta_{13}$, and $\theta_{23}$ are the three neutrino mixing angles, $\delta_{\rm CP}$ is the $CP$-violating Dirac phase, and $\alpha_{1,2}$ are two $CP$-violating Majorana phases. Experimental data, mainly from neutrino oscillation measurements, gives access to the differences of the neutrino masses squared as well as to the mixing angles and the Dirac phase \cite{NuFit2018, NuFit2019, NuFit2020}. The extracted mass squared differences directly translate into intervals for the two remaining masses as given in Tab.\ \ref{Tab:NeutrinoParameters}. We also show the intervals for the mixing angles and the Dirac phase. Note that the Majorana phases remain unconstrained by present data.

Following the Casas-Ibarra parametrization \cite{CasasIbarra2001}, the couplings $g_F$ and $g_{\Psi}$, which are responsible for the generation of the neutrino masses, are related to the parameters entering Eq.\ \eqref{Eq:NeutrinoDiagonalization} through
\begin{equation}
    {\cal G} ~=~ U_L \, D^{-1/2}_L \, R \, D_{\nu}^{1/2} \, U^{*}_{\rm PMNS} \,.
    \label{Eq:CasasIbarra}
\end{equation}
Here, $D_{\nu}$ is a diagonal matrix containing the neutrino mass eigenvalues and $U_{\rm PMNS}$ is the PMNS matrix. The matrix $D_L$ contains the eigenvalues of the matrix $M_L$ and $U_L$ is the associated rotation matrix. $D_L$ and $U_L$ only depend on the parameters related to the scalar and fermion sectors. A detailed discussion of the neutrino mass matrix and the Casas-Ibarra parametrization in the model under consideration is given in Apps.\ \ref{App:NeutrinoMassMatrix} and \ref{App:CasasIbarra}.

It is important to note that Eq.\ \eqref{Eq:CasasIbarra} is based on tree-level relations. In particular, the diagonalization of the scalar and fermion sectors entering the matrices $D_L$ and $U_L$ is done at the tree-level at this stage. Consequently, the neutrino masses will be affected when computing the scalar and fermion masses including one-loop corrections as it is done in our analysis using {\tt SPheno} instead of relying on tree-level relations as it is the case in Eq.\ \eqref{Eq:CasasIbarra}. 

\begin{figure}
    \centering
    \includegraphics[width=0.49\textwidth]{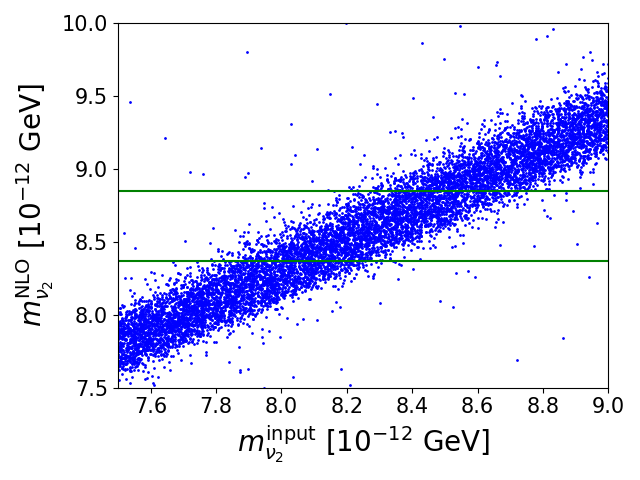}~~~
    \includegraphics[width=0.49\textwidth]{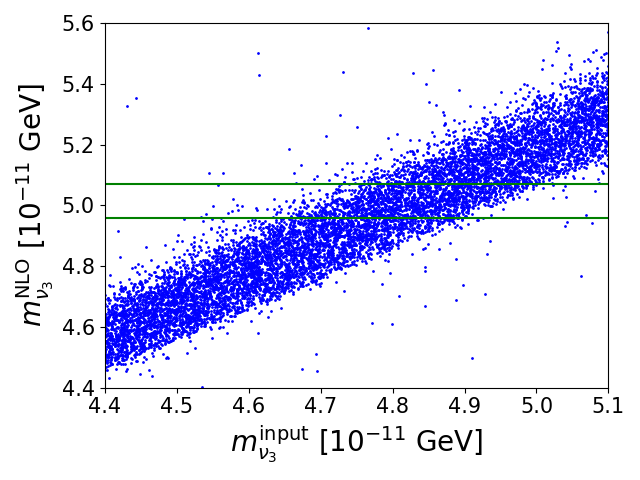}
    \vspace*{-8mm}
    \caption{Impact of one-loop corrections to scalar and fermion masses on the neutrino masses. We show the leading order (input) neutrino mass entering the Casas-Ibarra prescription against the mass obtained from {\tt SPheno} when taking into account one-loop corrections to the scalar and fermion masses as discussed in Secs.\ \ref{Sec:ScalarMasses} and \ref{Sec:FermionMasses}.}
    \label{Fig:NeutrinoMass}
\end{figure}

The impact of the one-loop corrections to the scalar and fermion masses on the neutrino mass calculation is illustrated in Fig.\ \ref{Fig:NeutrinoMass}. As can be seen, the impact is such that it shifts the neutrino masses towards higher values. Consequently, in the following study, we will choose the input neutrino mass values in a slightly shifted interval such that the resulting masses fall within the experimentally allowed ranges. 

% ======================================================
\section{Constraints and computational setup}
\label{Sec:Setup}
% ======================================================

In order to efficiently explore the parameter space of the model, we employ a Markov Chain Monte Carlo \cite{Markov1971} scanning technique based on the Metropolis-Hastings algorithm \cite{Metropolis1953, Hastings1970}. In an iterative procedure, the parameter space exploration is conditioned by a large number of (mostly experimental) constraints, implemented through the computation of the likelihood associated to a chosen parameter set. A crucial aspect of this technique is the comparison of prior and posterior parameter distributions, respectively obtained without and with imposing the constraints.

The Markov Chain Monte Carlo technique has proven to be a very powerful technique, especially in the exploration of high-dimensional parameter spaces. Originating from cosmology \cite{Trotta:2008qt} and astronomy \cite{Sharma:2017wfu}, this method has also been applied in phenomenological studies of supersymmetric extensions of the Standard Model \cite{Baer:2008jn, DeCausmaecker:2015yca} and the determination of parton distribution functions \cite{Mangin-Brinet:2017vkg}. To our knowledge, the present analysis is the first extensive study of a neutrino-mass model employing the Markov Chain Monte Carlo method.

In the following, we detail the computational details, the corresponding input parameters, as well as the imposed constraints.

% ======================================================
\subsection{Input parameters}
\label{Sec:Parameters}
% ======================================================

The model under consideration is determined by the parameters presented in Sec.\ \ref{Sec:Model} and summarized in Table \ref{Tab:T12AParameters}. Note that the parameter $\lambda_H$, although related to the Standard Model Higgs doublet and the experimentally known mass of the Higgs boson, is varied since one-loop contributions including the scalar additional fields may alter the prediction of the Higgs mass (see Sec.\ \ref{Sec:ScalarMasses}). Let us recall that the couplings $g_F$ and $g_{\Psi}$ related to the neutrino masses are computed from experimental neutrino data displayed in Table \ref{Tab:NeutrinoParameters} thanks to the Casas-Ibarra parametrization discussed in Sec.\ \ref{Sec:NeutrinoMasses}. 

\begin{table}
    \centering
    \renewcommand{\arraystretch}{1.2}
    \begin{tabular}{|c|c|}
        \hline
        Parameter & Interval \\
        \hline
        \hline
         $\lambda_H$ & $[0.1; 0.4]$  \\
        \hline
        \hline
         $M_S^2$, $M_{\Phi}^2$ & ~~$[0.5\cdot 10^6; 4\cdot 10^6]$~~ \\
        \hline
        \hline
         $T$ & $[-1000; 1000]$  \\
        \hline
        \hline
         $\lambda_{\Phi}$, $\lambda'_{\Phi}$, $\lambda''_{\Phi}$ & $[-2; 2]$  \\
        \hline
         $\lambda_S$, $\lambda_{4S}$, $\lambda_{4\Phi}$ & $[-2; 2]$  \\
        \hline
    \end{tabular} \qquad\quad
    \begin{tabular}{|c|c|}
        \hline
        Parameter & Interval \\
        \hline
        \hline
         $M_F$, $M_{\Psi}$ & ~~~$[700; 2000]$~~~   \\
        \hline
        \hline
         $y_1$, $y_2$ & $[-1.2; 1.2]$ \\
        \hline
        \hline
         $g_R^i$ ($i=e,\mu$) & $[-1.2; 1.2]$ \\
        \hline
         $g_R^{\tau}$  & $[-2.0; 2.0]$ \\
        \hline
        \hline
        $r$  & $[-1; 1]$  \\
        \hline
    \end{tabular}
    \caption{Independent input parameters of our Markov Chain Monte Carlo scan. All dimensionful quantities are given in GeV.}
    \label{Tab:T12AParameters}
\end{table}

\begin{table}
    \centering
    \renewcommand{\arraystretch}{1.2}
    \begin{tabular}{|c|c|}
        \hline
         Parameter & Interval \\
        \hline
        \hline
         $\Delta m_{12}^2$ & $[7.0 ; 7.84] \cdot 10^{-23}$ \\
        \hline
         $m_{\nu_2}$ & $[8.367; 8.854] \cdot 10^{-12}$ \\
        \hline
        \hline
         $\Delta m_{13}^2$ & $[2.47 ; 2.57] \cdot 10^{-21}$ \\
        \hline
         $m_{\nu_3}$ & $[4.96; 5.07] \cdot 10^{-11}$  \\
        \hline
    \end{tabular} ~\qquad~
    \begin{tabular}{|c|c|}
        \hline
         Parameter & Interval \\
        \hline
        \hline
         $\theta_{12}$ & $[31.90; 34.98]$ \\
        \hline
         $\theta_{13}$ & $[8.33; 8.81]$ \\
        \hline
         $\theta_{23}$ & $[46.8; 51.6]$ \\
        \hline
        \hline
         $\delta_{\rm CP}$ & $[143; 251]$ \\
        \hline
%       \hline
%         $\alpha_i$ ($i=1,2$) & $[0; 180[$  \\
%        \hline
    \end{tabular}
    \caption{Intervals at the  $2 \sigma$ confidence level for the neutrino masses and mixing parameters extracted from global fits of experimental neutrino data \cite{NuFit2020}. The neutrino masses are given in GeV, the angles and phases entering the PMNS matrix are given in degrees.}
    \label{Tab:NeutrinoParameters}
\end{table}

The intervals associated to each parameter have been determined using preliminary studies and are indicated in Tables \ref{Tab:T12AParameters} and \ref{Tab:NeutrinoParameters}. The Standard Model parameters are fixed to $m_{\rm top}^{\rm pole} = 173.5$ GeV, $m_b(m_b) = 4.18$ GeV, $m_{\tau} = 1.77669$ GeV, $1/\alpha_{\rm EW} = 127.9320$, $G_F = 1.166370 \cdot 10^{-5}$ GeV$^{-2}$, $\alpha_s(m_Z) = 0.1187$, $m_Z = 91.1887$ GeV \cite{PDG2020}.

% ======================================================
\subsection{Constraints and likelihood}
\label{Sec:Constraints}
% ======================================================

As a first constraint, we require a phenomenologically viable mass spectrum, i.e.\ no tachyonic states, and the lightest $\mathbb{Z}_2$-odd particle to be electrically neutral, such that it provides a viable dark matter candidate. Recall that the viable dark matter candidates within the model under consideration are $\chi^0_1$, $\phi^0_1$, and $A^0$. 

We then impose a number of constraints related to low-energy, flavour violating, and $CP$-violating observables. In particular, we impose experimental constraints from lepton-flavour violating processes such as $\ell_i \to \ell_j \gamma$ and $\ell_i \to \ell_j \ell_j \ell_k$ ($i,j,k = e,\mu,\tau$), muon-electron conversion rates in specific nuclei, and flavour-violating decays of the $Z$-boson. Moreover, we require the dark matter candidate to meet the relic density determined by the {\tt Planck} mission \cite{Planck2015}, allowing for a 2\% theoretical uncertainty in addition to the experimental error, and the direct detection cross-section not to exceed the limits published by the {\tt XENON1T} experiment \cite{XENONnT2020}. Finally, we require the Higgs-boson mass to match the experimental value determined from combined measurements of the {\tt ATLAS} and {\tt CMS} experiments \cite{Higgs2015} allowing for a 3 GeV uncertainty dominated by the theory error on the mass calculation.

\begin{table}
    \centering
    \renewcommand{\arraystretch}{1.2}
    \begin{tabular}{|c|c|}
        \hline 
         Observable & Constraint \\
        \hline
        \hline
         $m_H$ & $125.1 \pm 3.0$ GeV  \\
        \hline    
        \hline
         $\Omega_{\rm CDM}h^2$ & $0.1198 \pm 0.0042$ \\
        \hline
        \hline
         BR($\mu^- \to e^- \gamma$) & $< 4.2 \cdot 10^{-13}$  \\
        \hline
         BR($\tau^- \to e^- \gamma$) & $< 3.3 \cdot 10^{-8}$ \\
        \hline
         BR($\tau^- \to \mu^- \gamma$) & $< 4.4 \cdot 10^{-8}$ \\
        \hline
        \hline
         BR($\mu^- \to e^- e^+ e^-$) & $< 1.0 \cdot 10^{-12}$  \\
        \hline
         BR($\tau^- \to e^- e^+ e^-$) & $< 2.7 \cdot 10^{-8}$  \\
        \hline
         BR($\tau^- \to \mu^- \mu^+ \mu^-$) & $< 2.1 \cdot 10^{-8}$  \\
        \hline
         BR($\tau^- \to \mu^+ e^- e^-$) & $< 1.5 \cdot 10^{-8}$ \\
        \hline
         BR($\tau^- \to \mu^- e^+ e^-$) & $< 2.1 \cdot 10^{-8}$  \\
        \hline
         BR($\tau^- \to e^+ \mu^- \mu^-$) & $< 1.7 \cdot 10^{-8}$ \\
        \hline
         BR($\tau^- \to e^- \mu^+ \mu^-$) & $< 2.7 \cdot 10^{-8}$ \\
        \hline
    \end{tabular} \qquad
    \begin{tabular}{|c|c|}
        \hline 
         Observable & Constraint \\
        \hline
        \hline
         BR($Z^0 \to e^{\pm} \mu^{\mp}$) & $< 7.5 \cdot 10^{-7}$ \\
        \hline
         BR($Z^0 \to e^{\pm} \tau^{\mp} $) & $< 9.8 \cdot 10^{-6}$ \\
        \hline
         BR($Z^0 \to \mu^{\pm} \tau^{\mp} $) & $< 1.2 \cdot 10^{-5}$ \\
        \hline
        \hline
         BR($\tau^- \to e^- \pi^0 $) & $< 8.0 \cdot 10^{-8}$ \\
        \hline
         BR($\tau^- \to \mu^- \pi^0 $) & $< 1.1 \cdot 10^{-7}$ \\
        \hline
        \hline
         BR($\tau^- \to e^- \eta $) & $< 9.3 \cdot 10^{-8}$ \\
        \hline
         BR($\tau^- \to e^- \eta' $) & $< 1.6 \cdot 10^{-7}$ \\
        \hline
         BR($\tau^- \to \mu^- \eta $) & $< 6.5 \cdot 10^{-8}$ \\
        \hline
         BR($\tau^- \to \mu^- \eta' $) & $< 1.3 \cdot 10^{-7}$ \\
        \hline
        \hline
         ${\rm CR}_{\mu\to e}$(Ti) & $< 4.3 \cdot 10^{-12}$ \\
        \hline
         ${\rm CR}_{\mu\to e}$(Pb) & $< 4.6 \cdot 10^{-11}$ \\
        \hline
         ${\rm CR}_{\mu\to e}$(Au) & $< 7.0 \cdot 10^{-13}$ \\
    %    \hline
    %    \hline
    %     EDM_e & 1.1 \pm \. 0.11 \. $10^{-29}$ \\
    %    \hline
    %     EDM_{\mu} & 1.9 \pm \. 0.19 \. $10^{-29}$ \\
    %    \hline
    %     EDM_{\tau} & [-5.2;2.9] \. $10^{-16}$ &  \\
        \hline
        \end{tabular}
    \caption{Constraints stemming from Higgs mass measurement \cite{Higgs2015}, dark matter relic density \cite{Planck2018}, as well as flavour and low-energy precision data \cite{PDG2020}. In our setup, intervals given at the $1 \sigma$ confidence level are implemented using a Gaussian function, while upper limits are given at the 90\% confidence level and are implemented with a single-sided Gaussian allowing for a 10\% uncertainty.}
    \label{Tab:Constraints}
\end{table}

For a given parameter set $\vec{\theta}_n$, the adequacy with respect to the imposed constraints is expressed in terms of the likelihood ${\cal L}_n$. Assuming a Gaussian likelihood of uncorrelated observables, the latter is computed as the product of individual likelihoods with respect to each of the imposed constraints,
\begin{equation}
    \mathcal{L}_n ~\equiv~ \mathcal{L}(\Vec{\theta}_n, \Vec{O}) 
        ~=~ \prod_{i} \mathcal{L}_i(\Vec{\theta}_n,O_i) \,.
    \label{Eq:Likelihood}
\end{equation}
Here, the index $i$ runs over the various constraints, and $\vec{O}$ is the set of the observables.

In case of a measured observable, such as the Higgs-boson mass, $m_H$, or the dark matter relic density, $\Omega_{\rm CDM}h^2$, experimental intervals have been determined. Consequently, these constraints are implemented assuming a Gaussian profile with the given uncertainty $\sigma_i$. More precisely, the likelihood is computed as
\begin{equation}
    \ln {\cal L}_i(\vec{\theta_n}, O_i) 
        ~=~ -\frac{(O_i(\vec{\theta}_n)-O_i^{\rm exp})^2}{2\sigma_i^2} \,,
    \label{Eq:IndividualLikelihood}
\end{equation}
where $O_i(\vec{\theta}_n)$ denotes the predicted value of a given observable, $O_i^{\rm exp}$ is the experimental mean value, and $\sigma_i$ is the uncertainty associated to the observable $O_i$. Note that the latter may include experimental and theoretical uncertainties.

For the remaining constraints related to lepton-flavour violating processes as well as the dark matter direct detection, only upper limits have been derived by the various experiments. In practice, the corresponding likelihood computation is implemented as a step function, which is smeared as single-sided Gaussian with a width of 10\% of the value of the upper limit. The likelihood is ${\cal L}_i(\vec{\theta_n}, O_i) \equiv 1$ if the predicted value is below the experimental limit. Otherwise, the likelihood is calculated according to Eq.\ \eqref{Eq:IndividualLikelihood} with $O_i^{\rm exp}$ being the upper limit and $\sigma_i = 0.1 \cdot O_i^{\rm exp}$.

Concerning the neutrino sector, thanks to the Casas-Ibarra parametrization discussed in Sec.\ \ref{Sec:Parameters}, by construction all our parameter points should fulfill the constraints stemming from the neutrino sector. In practice, however, as the Casas-Ibarra procedure is implemented using the tree-level masses, while our final spectrum and neutrino mass calculation from {\tt SPheno} contains one-loop corrections, slight changes in the neutrino masses and the elements of the PMNS matrix may appear. For each point, we check again the compliance of the neutrino sector output of {\tt SPheno} against the experimental values given in Tab.\ \ref{Tab:NeutrinoParameters}.

Let us finally mention that we also impose the experimental constraints stemming from lepton electric dipole moments, namely ${\rm EDM}_e < 1.1 \cdot 10^{-29}$ and ${\rm EDM}_{\mu} < 1.9 \cdot 10^{-29}$ \cite{PDG2020}. However, these observables have no constraining power in the T1-2A framework under consideration.

% ======================================================
\subsection{Metropolis-Hastings algorithm}
\label{Sec:MCMC}

This algorithm aims at randomly exploring the parameter space while increasing the likelihood $\mathcal{L}_n$ in an iterative manner. Each Markov chain starts from a parameter point, which is randomly chosen within the intervals given in Tables \ref{Tab:NeutrinoParameters} and \ref{Tab:T12AParameters} such that its likelihood is (numerically) different from zero. Then, in each iteration, random values for the parameters $\vec{\theta}$ are picked in the vicinity of the previous accepted parameter point. In our study, the new proposed parameter value is computed according to
\begin{equation}
    \theta_{i}^{n+1} = \Pi\left\{ \theta_{i}^{n},\, \kappa \left( \theta_{i}^{\text{max}} - \theta_{i}^{\text{min}}\right) \right\} \,,
\end{equation}
where $\Pi \left\{ a,b \right\}$ is a Gaussian distribution with mean value $a$ and standard deviation $b$. The parameter $\kappa$ parametrizes the allowed jump length between two iterations, its value is chosen empirically in order to maximize the efficiency of the algorithm. Note that in case that the calculated value exceeds the limits of the corresponding interval (see Tables \ref{Tab:NeutrinoParameters} and \ref{Tab:T12AParameters}), the point is rejected.

For each chosen parameter set, the likelihood $\mathcal{L}^{n+1}$ is evaluated according to Eq.\ \eqref{Eq:Likelihood} and compared to the likelihood of the previous iteration $\mathcal{L}^{n}$. If $\mathcal{L}^{n+1} > \mathcal{L}^{n}$, the point is accepted and the chain will continue from this point. Otherwise, the new point is accepted with probability 
\begin{equation}
    p ~=~ \mathcal{L}^{n+1}/\mathcal{L}^n \,.
\end{equation}
If the point is rejected, a new random point is chosen in the vicinity of point $n$. Within this framework, the algorithm can move across larger regions while still converging to highest likelihood places. 

In typical chains, the likelihood value is supposed to grow significantly before converging towards a plateau near the global maximum. Because low-likelihood points generated before the convergence occurs are less relevant, we remove the first 50 points of the chain (burn-in length). This parameter is also set empirically by observing the likelihood curve across the chains.

In a high-dimensional parameter space (in the present case: 26 parameters), the quality of the exploration relies more on the total number of chains than the individual lengths of the chains themselves. Note that different starting points, chosen randomly, may lead to different likelihood maximums. For the following analysis, we have generated a total of 86.100 accepted parameter points stemming from 246 independent Markov chains.

% ======================================================
\subsection{General computational setup}
\label{Sec:Computational}
% ======================================================

Starting from the input parameters given in the $\overline{\rm MS}$-scheme at the scale $Q = m_t^{\rm pole}$, for each parameter point, we compute the physical mass spectrum at the one-loop level employing the numerical program {\tt SPheno 4.0.4} \cite{SPheno2003, SPheno2012}, where we have implemented the model T1-2A making use of the {\tt Mathematica} package {\tt SARAH 4.3.14} \cite{SARAH2010, SARAH2011, SARAH2013, SARAH2014}. This setup is also used to compute the observables related to lepton-flavour violation and low-energy measurements, which will be detailed in Sec.\ \ref{Sec:Constraints}. The dark matter relic density and the direct detection cross-sections are numerically evaluated using {\tt micrOMEGAs 5.0.8} \cite{MO2001, MO2004, MO2007a, MO2007b, MO2013, MO2018}, where again the corresponding {\tt CalcHEP} \cite{CalcHep_Belyaev:2012qa} model files have been generated using {\tt SARAH}. The parameter values are communicated between the two codes using the {\tt SLHA} file format \cite{SLHA1, SLHA2}.
% ======================================================
\section{Results}
\label{Sec:Results}
% ======================================================

In this Section, we will present the main results from our Markov Chain Monte Carlo studies presented in Sec.\ \ref{Sec:Setup}. After a discussion of the couplings and the constraints from the lepton and neutrino sectors, we will in particular focus on the resulting dark matter phenomenology.

% ======================================================
\subsection{Couplings, neutrino masses, and lepton flavour violation}
% ======================================================

\begin{figure}
    \centering
    \includegraphics[width=0.49\textwidth]{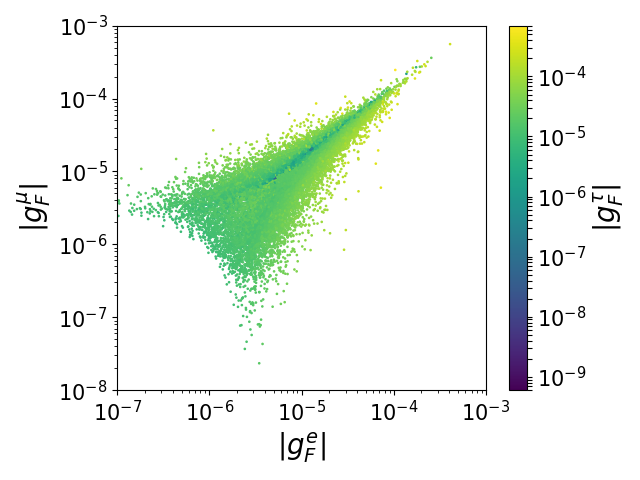}~~~
    \includegraphics[width=0.49\textwidth]{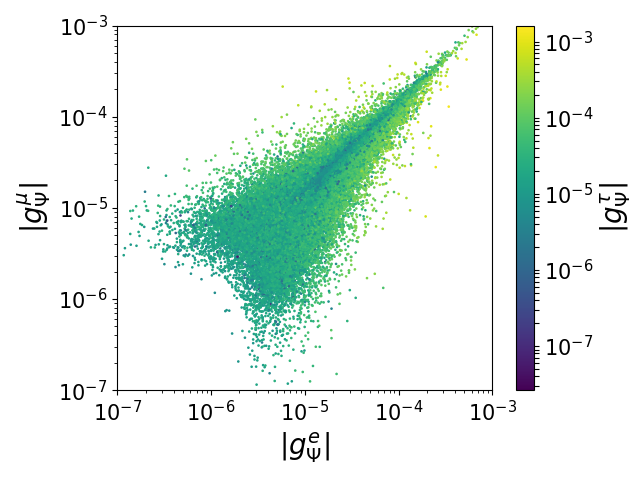}
    \vspace*{-8mm}
    \caption{Distribution of the absolute values of the coupling parameters $g_F^i$ and $g_{\Psi}^i$ ($i=e,\mu,\tau$) obtained from the Markov Chain Monte Carlo analysis.}
    \label{Fig:CouplingsFPsi}
\end{figure}

The couplings $g_F$ and $g_{\Psi}$ are strongly related to the measured neutrino masses and mixing parameters through Eq.\ \eqref{Eq:CasasIbarra}. The obtained distributions of the absolute values are shown in Fig.\ \ref{Fig:CouplingsFPsi}. As can be seen, the couplings are constrained to be rather small and cannot exceed about $10^{-3}$. Consequently, the couplings $g_F$ and $g_{\Psi}$ cannot be expected to contribute significantly to lepton-flavour violating processes such as $\mu \to e\gamma$ and other decays. In other words, the lepton-flavour violating processes have almost no impact on the viable values of the couplings $g_F$ and $g_{\Psi}$.

\begin{figure}
    \centering
    ~~\includegraphics[width=0.54\textwidth]{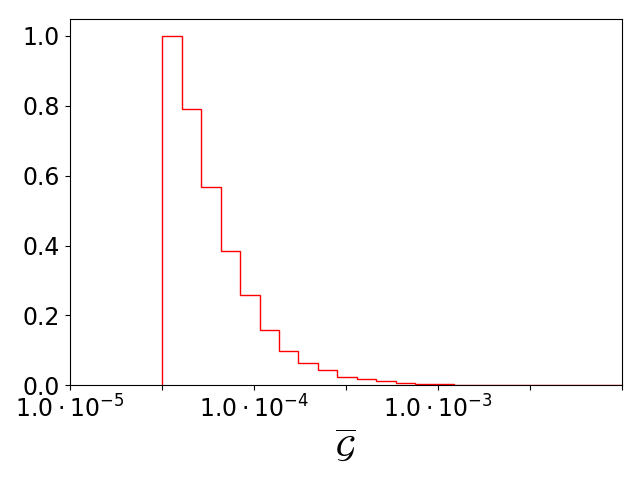}
    \vspace*{-5mm}
    \caption{Distribution of the geometric mean value $\overline{\cal G}$ of the coupling elements defined in Eq.\ \eqref{Eq:MatrixNorm}.}
    \label{Fig:CouplingsMean}
\end{figure}

As complementary information, we compute the geometrical mean value of the elements of the matrix ${\cal G}$ (regrouping the couplings $g_F$ and $g_{\Psi}$, see Eq.\ \eqref{Eq:CouplingMatrix}),
\begin{equation}
    \overline{{\cal G}} ~=~ \left( g_F^e \, g_F^{\mu} \, g_F^{\tau} \, g_{\Psi}^e \, g_{\Psi}^{\mu} \, g_{\Psi}^{\tau} \right)^{1/6} \,.
    \label{Eq:MatrixNorm}
\end{equation}
The numerical values of this parameter are shown in Fig.\ \ref{Fig:CouplingsMean}. We see that the geometrical mean value of the couplings is relatively stable peaking around $\overline{\cal G} \approx 3.2 \cdot 10^{-5}$. This value represents the typical order of magnitude needed to meet the experimental constraints from the neutrino sector. If one entry of the matrix ${\cal G}$ is numerically larger (smaller), it has to be compensated by other elements which are then smaller (larger). Note that the width of the shown distribution is caused by the fact that we vary the parameters of the scalar and fermion sectors appearing in the matrix $M_L$ of Eq.\ \eqref{Eq:CasasIbarra}.

\begin{figure}
    \centering
    \includegraphics[width=0.49\textwidth]{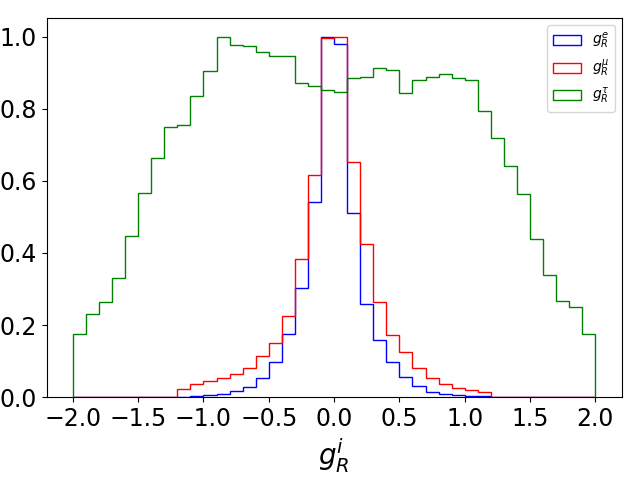}
    \vspace*{-2mm}
    \caption{Distributions of the couplings $g_R^i$ ($i=e,\mu,\tau$) obtained from the Markov Chain Monte Carlo analysis.}
    \label{Fig:CouplingsR}
\end{figure}

Let us now discuss the coupling $g_R$, which is not related to the neutrino sector, but constrained from lepton flavour data. Fig.\ \ref{Fig:CouplingsR} shows the distributions that we obtain for the three individual couplings after imposing the constraints in our Markov Chain Monte Carlo study. The numerical values of $g_R^e$ and $g_R^{\mu}$ are bound to a moderately large interval due to the high experimental precision in the  searches for $\mu$-$e$ transitions. Searches involving tau leptons do not meet this level of precision (see Table \ref{Tab:Constraints}) and, consequently, the corresponding coupling parameter $g_R^{\tau}$ is less constrained, allowing values of up to $|g_R^{\tau}| \lesssim 2.0$. The localized peaks around $|g_R^{\tau}| \sim 1.0$ is explained by the interplay of the relic density constraint and the flavour-violating tau decays. The latter forbid too high values of $|g_R^{\tau}|$, while non-zero couplings increase the relative contribution of dark matter (co-)annihilation into tau final states (see Sec.\ \ref{Sec:DarkMatter}).

% ======================================================
\subsection{Dark matter mass and nature}
\label{Sec:DarkMatter}
% ======================================================

Coming to the second motivation of the present model after providing neutrino masses, we will now discuss the dark matter phenomenology of the framework after including all constraints discussed in Sec.\ \ref{Sec:Constraints}. Recall that the present model includes three viable dark matter candidates, the lightest neutral fermion $\chi^0_1$, the lightest neutral scalar $\phi^0_1$, and the pseudo-scalar $A^0$.

\begin{figure}
    \centering
    \includegraphics[width=0.49\textwidth]{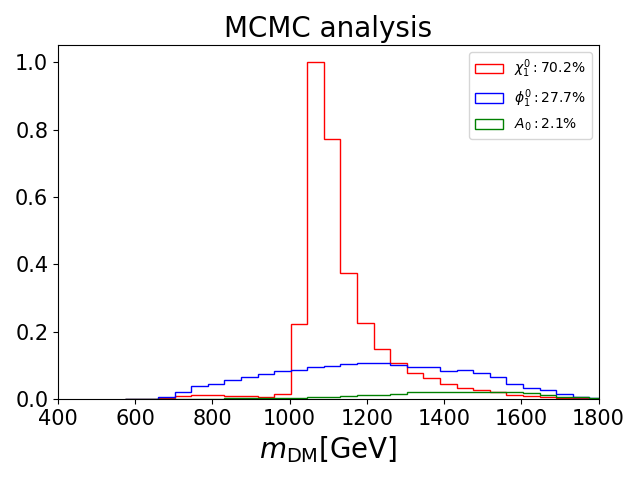}~~~
    \includegraphics[width=0.49\textwidth]{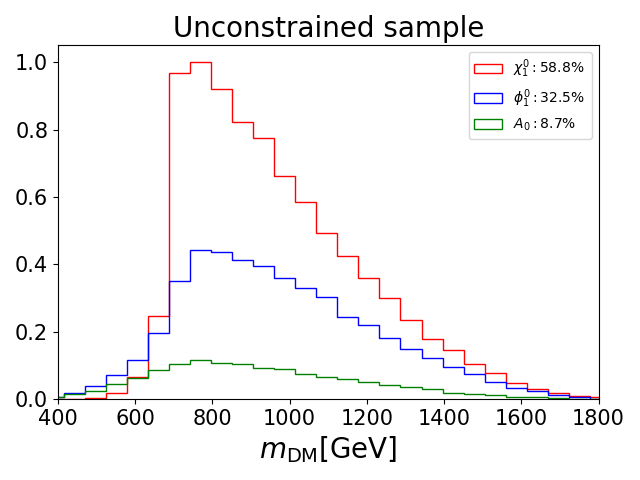}
    \vspace*{-8mm}
    \caption{Distribution of the dark matter mass obtained from the Markov Chain Monte Carlo study imposing the constraints of Tab.\ \ref{Tab:Constraints} (left) and for a random scan using the same parameter intervals but without imposing the constraints (right). The three histograms show the distribution for the three possible dark matter candidates ($\chi^0_1$, $\phi^0_1$, $A^0$). The percentages in the legend correspond to the relative occurrence of each dark matter candidate.}
    \label{Fig:MassDM}
\end{figure}

We start by investigating the mass interval that can be expected for the dark matter particle. In Fig.\ \ref{Fig:MassDM} we show the obtained distribution of the dark matter mass together with its respective nature. Moreover, we compare the situation after imposing the constraints of Table \ref{Tab:Constraints}, i.e.\ the viable parameter points obtained from the Markov Chain Monte Carlo (MCMC) analysis, to the situation without imposing these constraints, obtained from a pure random scan with the only requirement that the dark matter particle is electrically neutral. The unconstrained sample aims at knowing the prior distribution of the dark matter mass. Comparing to the posterior distribution (obtained from the MCMC analysis) allows to conclude concerning the impact of the constraints on the dark matter mass distribution.

First, we observe that the model intrinsically prefers fermionic dark matter over scalar or pseudo-scalar dark matter. Already in the case without imposing constraints, the former accounts for about 60\% of the accepted parameter points. Imposing the experimental constraints, this percentage is slightly increased to about 70\%. At the same time, the percentage of pseudo-scalar dark matter decreases, mainly due to the dark matter direct detection constraint, which will be discussed in more detail in Sec.\ \ref{Sec:DirectDetection}. 

Second, after imposing the constraints, the mass distribution associated to fermionic dark matter displays a rather sharp shape peaking at $m_{\chi^0_1} \sim 1.1$ TeV. In contrast, the scalar and pseudo-scalar dark matter feature much wider distributions reaching from about 700 GeV to about 1700 GeV. The case of scalar dark matter has been discussed in detail in Ref.\ \cite{Esch2018}.

\begin{figure}
    \centering
    \includegraphics[width=0.49\textwidth]{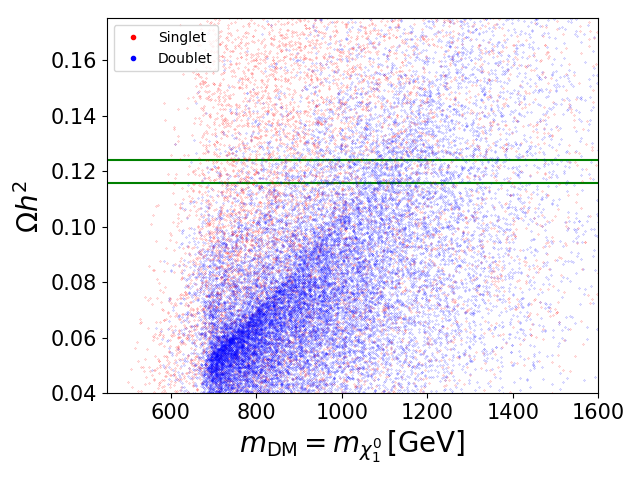}~~~
    \includegraphics[width=0.49\textwidth]{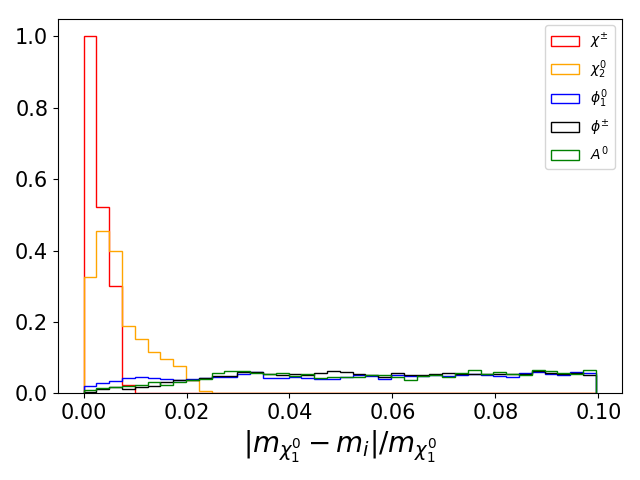}
    \vspace*{-8mm}
    \caption{Left: Correlation of the fermion dark matter relic density and the fermion dark matter mass obtained from a random scan without imposing the experimental constraints of Tab.\ \ref{Tab:Constraints}. Blue points correspond to doublet-dominated fermions, while red points indicate singlet-dominated configurations. The horizontal lines are the Planck limits for the relic density. Right: Mass difference between the heavier neutral fermion, charged fermion, the lightest scalar, charged scalar, and the pseudo-scalar with respect to the lightest fermion mass for the case where the lightest fermion is the dark matter candidate using the MCMC sample.}
    \label{Fig:DarkMatterPheno} 
\end{figure}

The striking behaviour in the fermionic case can be traced to requiring the predicted relic density to meet the Planck limit of $\Omega_{\rm CDM}h^2 \approx 0.12$ (see Table \ref{Tab:Constraints}). In order to understand the exact mechanism, we show in Fig.\ \ref{Fig:DarkMatterPheno} the predicted relic density projected on the lightest fermion mass together with information about the singlet-doublet composition of the lightest fermion. The singlet-dominated (doublet-dominated) parameter points have a singlet component of more than 90\% (less than 10\%). In total, about one third of the points is singlet-dominated, while about two thirds are doublet-dominated. Moreover, we show in Fig.\ \ref{Fig:DarkMatterPheno} the relative mass differences between the lightest fermion and the remaining $\mathbb{Z}_2$-odd particles, which are potential co-annihilation partners. 

In the left part of Fig.\ \ref{Fig:DarkMatterPheno} it can be seen that the parameter configurations are about uniformly distributed in the case of singlet-like dark matter. In this case, pair-annihilations of the lightest fermion are dominant, while co-annihilations with the other particles can occur depending on the exact mass configurations of the individual parameter points. 

The situation is different for the case of doublet-dominated fermionic dark matter candidates. As can be seen in the left panel of Fig.\ \ref{Fig:DarkMatterPheno}, the corresponding parameter configurations are grouped in a rather specific region of the $\Omega_{\chi^0_1}h^2$--$m_{\chi^0_1}$ plane. This is explained by the fact that in the doublet case, the physical masses of the lightest, the second-lightest, and the charged fermions are very close. This manifests in the right panel of Fig.\ \ref{Fig:DarkMatterPheno} as the pronounced peak for precisely these states, meaning that indeed $m_{\chi^0_1} \approx m_{\chi^0_2} \approx m_{\chi^{\pm}}$. Consequently, co-annihilations between these states are dominant, and the relic density depends essentially on the overall mass scale. This explains the highly densely populated region of doublet-dominated points visible in Fig.\ \ref{Fig:DarkMatterPheno}. As can be seen, following this region towards higher masses, the experimentally observed relic density is achieved for masses of about 1.0 -- 1.2 TeV, corresponding to the peak observed in the mass distribution shown in Fig.\ \ref{Fig:MassDM}. In this situation, the fermionic dark matter mainly (co-)annihilates into tau leptons and quarks, while in specific configurations also $W$- and $Z$-boson may occur in the final state. 

% ======================================================
\subsection{Dark matter direct detection}
\label{Sec:DirectDetection}
% ======================================================

\begin{figure}
    \centering
    \includegraphics[width=0.49\textwidth]{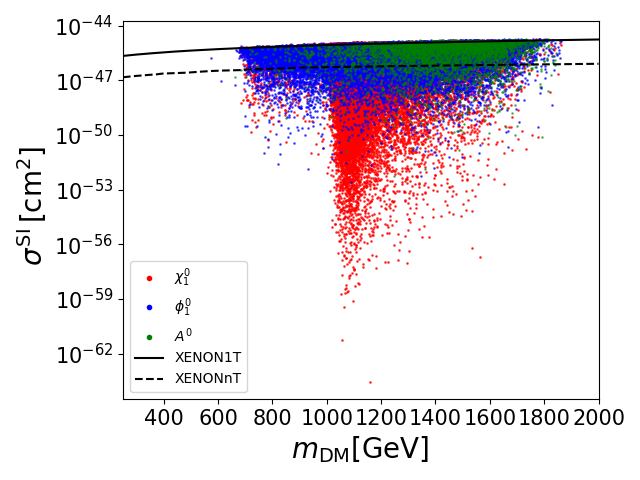}~~~
    \includegraphics[width=0.492\textwidth]{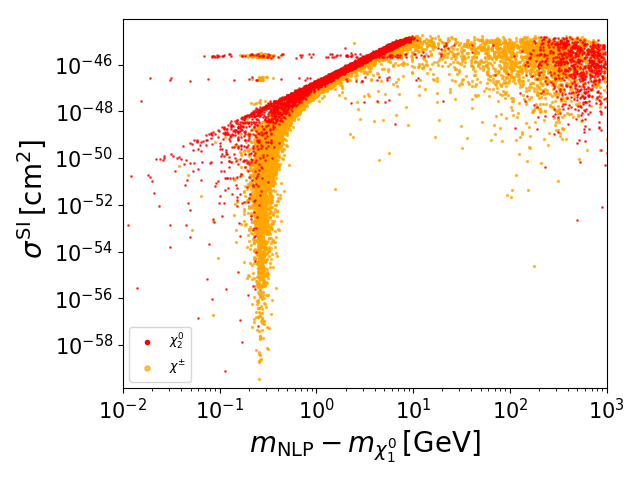}
    \vspace*{-8mm}
    \caption{Left: Spin-independent dark matter direct detection cross-section as a function of the dark matter mass. The colours correspond to the different natures of the dark matter. Right: Spin-independent dark matter direct detection cross-section versus the mass difference between the dark matter and the next-to-lightest particle (NLP) for the case of fermionic dark matter. The colours correspond to the different natures of the next-to-lightest particle.}
    \label{Fig:DMDD}
\end{figure}

Let us finally comment on the dark matter direct detection and its implications on the model parameter space. We show in the left part of Fig.\ \ref{Fig:DMDD} the spin-independent scattering cross-section of a dark matter particle off a nucleon as a function of the dark matter mass for each valid parameter point of our Markov Chain Monte Carlo analysis. Points not satisfying the current limits from the {\tt XENON1T} experiment \cite{XENON1T2018} are not shown. 

This limit excludes a large part of pseudo-scalar dark matter, the remaining valid configurations being situated relatively close to the exclusion limit. A large majority of these points lie within the reach of the {\tt XENONnT} experiment \cite{XENONnT2020}, meaning that the latter will be able to exclude pseudo-scalar, and thus doublet-dominated scalar dark matter, in a near future. The scalar dark matter configurations allowing smaller scattering cross-sections are mainly singlet-dominated. This corresponds to the case discussed in Ref.\ \cite{Esch2018}. 

Coming to fermionic dark matter, as discussed in Sec.\ \ref{Sec:DarkMatter}, the region that is most densely populated corresponds to the case of doublet-dominated dark matter. This scenario can feature rather small scattering cross-sections, which are well below the reach of future experiments. It is to be noted that the cross-section is correlated with the mass difference of the dark matter fermion and the next-to-lightest particle, as can be seen in the right part of Fig.\ \ref{Fig:DMDD}. More precisely, the points that are expected to evade the {\tt XENONnT} projected limits feature very small mass splittings within the fermionic doublet.

Consequently, it will be interesting to investigate further the associated phenomenology, e.g., concerning collider searches or related signatures. The case of singlet-dominated fermion dark matter can be excluded more easily.

Note that we have also included the limits from spin-independent direct dark matter searches leading to similar conclusions as the spin-dependent analysis presented above.
% ======================================================
\section{Summary and collider perspectives}
\label{Sec:Outlook}
% ======================================================

\begin{table}
    \centering
    \begin{tabular}{|c|c|}
        \hline
        Parameter & Interval \\
        \hline
        \hline
         $\lambda_H$ & $[0.1; 0.3]$  \\
        \hline
        \hline
         $T$ & $[-500; 500]$  \\
        \hline
        \hline
         $\lambda_S$ & $[-1.0; 1.0]$  \\
        \hline
         $\lambda_{\Phi}$ & ~~$[-1.25; 0.75]$~~ \\
        \hline
         $\lambda'_{\Phi}$ & $[-1.50; 1.00]$ \\
        \hline
         $\lambda''_{\Phi}$ & $[-1.75; 1.50]$ \\
        \hline
        \hline
         $\lambda_{4S}$, $\lambda_{4\Phi}$ & $[-2.0; 2.0]$  \\
        \hline
    \end{tabular} \qquad\quad
    \begin{tabular}{|c|c|}
        \hline
        Parameter & Interval \\
        \hline
        \hline
         $M_S^2$ & ~~$[0.5; 4.0] \cdot 10^6$~~ \\
        \hline
         $M_{\Phi}^2$ & ~~$[1.5; 4.0] \cdot 10^6$~~ \\
        \hline
        \hline
         $M_F$ & ~~~$[1300; 2000]$~~~   \\
        \hline
         $M_{\Psi}$ & $[1000; 1300]$   \\
        \hline
        \hline
         $y_1$, $y_2$ & $[-0.75; 0.75]$ \\
        \hline
        \hline
         $g_R^e$, $g_R^{\mu}$ & $[-0.25; 0.25]$ \\
        \hline
         $g_R^{\tau}$  & $[-1.5; 1.5]$ \\
        \hline
    \end{tabular}
    \caption{Intervals for the model parameters which can be expected to lead to viable parameter configurations obtained from the Markov Chain Monte Carlo analysis. All dimensionful parameters are given in GeV. Approximate viable intervals for the coupling parameters $g_F^i$ and $g_{\Psi}^i$ ($i=e,\mu,\tau$) can be extracted from Fig.\ \ref{Fig:CouplingsFPsi}.}
    \label{Tab:Results}
\end{table}

In summary, the experimental constraints listed in Table \ref{Tab:Constraints} leave important room for viable parameter points within the scotogenic T1-2A framework. As a general result of our Markov Chain Monte Carlo study, we present in Table \ref{Tab:Results} the parameter ranges which can be expected to lead to viable parameter configurations. Note that most intervals given in Table \ref{Tab:Results} are restricted with respect to the input intervals given in Table \ref{Tab:T12AParameters} due to the imposed constraints. 

It is also to be noted that, due to possible interferences between different parameters, not all configurations in the direct product of the given intervals will lead to phenomenologically viable parameter sets. We therefore recommend to use the values in Table \ref{Tab:Results} as a guideline, but check individual parameter configurations in view of the numerous constraints that apply to this kind of model. In the case where the couplings $g_F$ and $g_{\Psi}$ are chosen as free input parameters, the constraints from the neutrino sector are to be verified in addition. Let us recall that in our study the couplings computed from the Casas-Ibarra parametrization fulfill these constraints by construction (see Sec.\ \ref{Sec:NeutrinoMasses}).

The model under consideration includes for possible scenarios concerning the dark matter candidate. The lightest $\mathbb{Z}_2$-odd particle can be either a scalar or a fermion, and in both cases it can be singlet- or doublet-dominated. In Table \ref{Tab:Scenarios}, we indicate reference scenarios for each case, chosen among those presenting the highest likelihood values from our Markov Chain Monte Carlo analysis and having a rather typical mass configuration for the given scenario type. In the following, giving special focus to the case of doublet-like fermion dark matter, we briefly discuss the typical mass configurations together with the associated expected collider phenomenology. A detailed study of the related signatures is, however, beyond the scope of this work. 

The complete mass spectra in {\tt SLHA 2} format \cite{SLHA2} including the corresponding input parameters as well as the complete decay information can be found as ancillary files associated to the electronic submission of the present publication.

\begin{table}
    \centering
    \begin{tabular}{|c||cc|c||c|c|}
        \hline
         & \multicolumn{2}{c|}{~~~~Doublet fermion~~~~} & Singlet fermion & Doublet scalar & ~Singlet scalar~ \\
         & ~~~(I) & (II) & & & \\ 
        \hline
        \hline
        $m_{\chi^0_1}$   & ~~~1077.6 & 1062.2 &  956.2 & 1647.3 & 1373.2 \\
        $m_{\chi^0_2}$   & ~~~1081.9 & 1069.3 & 1130.0 & 1807.5 & 1785.5 \\
        $m_{\chi^0_3}$   & ~~~1776.1 & 1856.1 & 1133.3 & 1976.5 & 1808.6 \\
        $m_{\chi^{\pm}}$ & ~~~1080.1 & 1062.5 & 1117.8 & 1805.9 & 1782.8 \\
        \hline
        \hline
        $m_{\phi^0_1}$   & ~~~1381.1 & 1280.5 & 1358.2 & 1452.2 & 1165.3 \\
        $m_{\phi^0_2}$   & ~~~1637.4 & 1388.6 & 1664.4 & 1591.8 & 1751.8 \\
        $m_{A^0}$        & ~~~1388.1 & 1312.9 & 1381.6 & 1465.4 & 1764.7 \\
        $m_{\phi^{\pm}}$ & ~~~1394.1 & 1317.9 & 1360.8 & 1453.9 & 1765.8 \\
        \hline
    \end{tabular}
    \caption{Typical phenomenologically viable mass configurations in the T1-2A scotogenic model resulting from our Markov Chain Monte Carlo analysis. We indicate the physical masses of the new fermions and scalars for the five chosen dark matter configurations. All masses are given in GeV.}
    \label{Tab:Scenarios}
\end{table}

% ==========================================================
\subsection{Doublet fermion dark matter}
% ==========================================================

As already discussed in Sec.\ \ref{Sec:Results}, this case represents the majority of viable parameter points in the model under consideration, and also the majority of the parameter sets presenting the highest global likelihood values. Due to a common relatively small mass parameter $M_{\Psi}$, the second-lightest neutral fermion as well as the charged fermion are very close in mass to the lightest fermionic state, which is the dark matter candidate. The latter typically has a mass of about $m_{\chi^0_1} \sim 1050 - 1150$ GeV with mass differences around $(m_{\chi^0_2}-m_{\chi^0_1}) \sim (m_{\chi^{\pm}}-m_{\chi^0_1}) \sim 0.1 - 10$ GeV (see Fig.\ \ref{Fig:DMDD}).

As explained in Sec.\ \ref{Sec:Results}, this very small mass difference can be traced to the relic density constraint. More precisely, the Planck limit given in Tab.\ \ref{Tab:Constraints} can be met thanks to sizeable co-annihilation contributions. This drives the doublet nature of the fermionic dark matter in a majority of viable parameter points.

\begin{figure}
    \centering
    \includegraphics[scale=1.0]{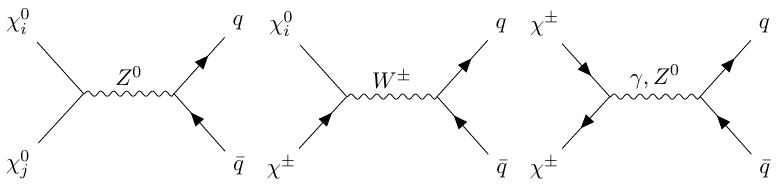}
    \caption{Dominant co-annihilation processes for the case of doublet-dominated fermionic dark matter. The indices $i,j=1,2$ run over the different neutral fermion states involved in the co-annihilation.}
    \label{Fig:DiagramCoannFermion}
\end{figure}

The lightest fermion $\chi^0_1$ then co-annihilates with the second-lightest neutral fermion $\chi^0_2$ or the charged fermion $\chi^{\pm}$, which are both doublet-like as well. Due to the very small mass gap and the stronger interaction cross-section, even $\chi^+-\chi^-$ annihilation may contribute significantly. The mass of the singlet-like third neutral fermion $\chi^0_3$ is not correlated to the other fermion masses in this situation. In the present model, the $\chi^0_1-\chi^0_1$ pair annihilation as well as the mentioned co-annihilations essentially lead to quark-antiquark final states. The corresponding Feynman diagrams are shown in the Fig.\ \ref{Fig:DiagramCoannFermion}. Additional diagrams with Higgs exchange are less prominent due to the doublet nature of the co-annihilating particles and the relatively small Yukawa couplings of the light quarks.

In typical configurations featuring doublet-dominated fermionic dark matter, the co-annihilation contributions are dominant, accounting for up to about 75\% of the total annihilation cross-section entering the relic density calculation. For the two example scenarios given in Table \ref{Tab:Scenarios}, which fulfill the relic density constraint of $\Omega_{\chi^0_1}h^2 \approx 0.12$, the would-be relic density when excluding co-annihilations is much higher, $\Omega^{\rm no-coann.}_{\chi^0_1}h^2 \approx 0.27$ for both example scenarios.

At the Large Hadron Collider (LHC), the production of $\chi^0_2$ and $\chi^{\pm}$ may proceed through $s$-channel vector or Higgs-boson exchange. The associated cross-sections are expected to be similar to those for neutralinos and charginos in the Minimal Supersymmetric Standard Model (MSSM).

Due to the very small mass gaps, only three-body decays are possible in this specific configuration. We expect the two particles to decay mainly into quark-antiquark pairs and missing energy, while decays into lepton or neutrino pairs and missing energy are expected to be less prominent. Note that smaller mass gaps generally allow for smaller dark matter direct detection cross-sections. For the example scenario (I) given in Table \ref{Tab:Scenarios} the second-lightest fermion mainly decays according to
\begin{align}
    \begin{split}
    {\rm BR}( \chi^0_2 \to \chi^0_1 d \bar{d} ) ~&\approx~ 0.27 \,, \\
    {\rm BR}( \chi^0_2 \to \chi^0_1 s \bar{s} ) ~&\approx~ 0.27 \,, \\
    {\rm BR}( \chi^0_2 \to \chi^0_1 u \bar{u} ) ~&\approx~ 0.21 \,, \\
    {\rm BR}( \chi^0_2 \to \chi^0_1 c \bar{c} ) ~&\approx~ 0.10 \,, 
    \end{split}
\end{align}
while the most important decay channels of the charged fermion are
\begin{align}
    \begin{split}
    {\rm BR}( \chi^- \to \chi^0_1 d \bar{u} ) ~&\approx~ 0.50 \,, \\
    {\rm BR}( \chi^- \to \chi^0_1 s \bar{c} ) ~&\approx~ 0.15 \,, \\
    {\rm BR}( \chi^- \to \chi^0_1 \mu^- \nu ) ~&\approx~ 0.16 \,, \\
    {\rm BR}( \chi^- \to \chi^0_1 e^- \nu )   ~&\approx~ 0.13 \,,
    \end{split}
\end{align}

where the notation $\nu$ comprises the three neutrino states. Here, we only indicate the decay channels with a branching ratio of at least 0.1. The relic density is achieved through co-annihilations between the three states $\chi^0_1$, $\chi^0_2$, and $\chi^{\pm}$ mainly into quark-antiquark pairs. The spin-independent direct detection cross-section is $\sigma^{\rm SI} \approx 1.0 \cdot 10^{-46}$ cm$^2$ and thus reachable by the {\tt XENONnT} experiment.

Choosing a parameter set with an even smaller mass gap allows to avoid the projected {\tt XENONnT} constraint (see Fig.\ \ref{Fig:DMDD}). Example scenario (II) of Table \ref{Tab:Scenarios} has been chosen in this spirit. Here, the singlet admixture is smaller than in the previous case and the physical masses are even closer. The relic density is still governed by the same co-annihilation channels as in the previous parameter point. However, the direct detection is much smaller, $\sigma^{\rm SI} \approx 7.0 \cdot 10^{-51}$ cm$^2$, well below the projected reach of ${\tt XENONnT}$. 

A further consequence of such small mass splitting concerns the decay of the fermions $\chi^0_2$ and $\chi^{\pm}$. The above decays are still open, but kinematically strongly suppressed. The dominant decay mode is then $\chi^0_2 \to \chi^0_1 \gamma$ mediated at the one-loop level. The associated decay width is of the order of $10^{-10}$ GeV, such that the decaying particle can be considered as long-lived. Similar conclusions may hold for the decay of the charged fermion.

% ==========================================================
\subsection{Singlet fermion dark matter}
% ==========================================================

In this case, the dark matter mass is not related to the other fermion masses any more. Consequently, for the heavier fermionic states a large variety of decay modes is possible, depending on the exact mass configuration including the potential intermediate scalars. The dark matter mass can be of $m_{\chi^0_1} \sim 800 - 1400$ GeV (see also Fig.\ \ref{Fig:DarkMatterPheno}). The dominant annihilation channels and final states depend on the exact configuration of the remaining $\mathbb{Z}_2$-odd particles.

% ==========================================================
\subsection{Doublet scalar dark matter}
% ==========================================================

In this case, the lightest scalar is close in mass to the pseudo-scalar and the charged scalar. Typical masses are in the range $m_{\phi^0_1} \sim 1100 - 1500$ GeV while $(m_{A^0}-m_{\phi^0_1}) \sim (m_{\phi^{\pm}}-m_{\phi^0_1}) \sim 0.1 - 10$ GeV, similar to the fermion case, while the singlet-like state $\phi^0_2$ can be significantly heavier. Production at the LHC may proceed via gauge or Higgs-boson exchange with cross-sections which are typical for two-Higgs doublet models. 

Here, dominant co-annihilations essentially lead to final states with gauge and Higgs bosons, e.g., $\phi^{\pm} \phi^0_1 \to Z W^{\pm}$. Again, the would-be relic density without co-annihilations would clearly exceed the Planck upper limit.

As no tree-level decays are allowed in this configuration, the produced pseudo-scalar $A^0$ will undergo a loop-mediated decay into the lightest scalar $\phi^0_1$ and a photon. For the example scenario given in Table \ref{Tab:Scenarios}, the associated decay width is around $4 \cdot 10^{-6}$ GeV, making the pseudo-scalar rather long-lived.

The situation is similar for the charged scalar $\phi^{\pm}$. No decay channels being available at the tree-level, this state will decay into the lightest scalar $\phi^0_1$ plus light quarks or lepton-neutrino pairs at the loop-level. Again, long-lived charged scalars can be expected in this case.

Let us recall that the case of doublet scalar dark matter can be expected to be challenged by upcoming direct dark matter searches (see Sec.\ \ref{Sec:DirectDetection}). 

% ==========================================================
\subsection{Singlet scalar dark matter}
% ==========================================================

Finally, in the singlet-scalar dark matter case, typical masses are $m_{\phi^0_1} \sim 1000$ GeV, the other scalars as well as the fermions may be significantly heavier. As in the fermionic singlet case, a variety of decay options is available for the heavier states, again depending on the exact parameter configuration. The dark matter phenomenology of this scenario has been studied in more detail in Ref.\ \cite{Esch2018}.

% ======================================================
\section{Conclusion}
\label{Sec:Conclusion}
% ======================================================

The ``T1-2A'' model is a rather general framework of the scotogenic type, presenting a very rich phenomenology. We have presented a complete analysis of the associated parameter space, taking into account constraints from the Higgs sector, the neutrino sector, lepton-flavour violating processes, and dark matter. To our knowledge, this is the first analysis of a scotogenic framework taking into account the complete parameter space together with such a large variety of constraints.

While neutrino data governs the couplings of the new particles to the left-handed leptons ($g_F$ and $g_{\Psi}$ in Eq.\ \eqref{Eq:LagCouplings}), the couplings to the right-handed leptons ($g_R$) are mainly constraint by the decays $\mu \to e \gamma $ and $\mu \to 3e$ as well as $\mu-e$ conversion rates in nuclei. The dark matter relic density and the direct detection constrain the mass of the new particles as well as their Yukawa couplings. The {\tt XENONnT} projection will have an impact on the dark matter nature by excluding a large part of parameter space featuring pseudo-scalar dark matter.

Focusing on dark matter phenomenology, contrary to previous publications focusing on one single dark matter candidate, we have considered all three possibilities, namely scalar, pseudo-scalar, and fermionic dark matter. The latter turns out to be preferred in view of the imposed constraints. 

Moreover, fermionic dark matter is expected to be doublet-dominated. Our analysis has shown that in this case, a rather precise mass range of about 1.0 -- 1.2 TeV can be expected for the dark matter fermion. In this parameter region, the correct dark matter relic density is achieved through co-annihilations with the other fermions, while direct detection constraints can be avoided. 

The elements discussed in the present paper may be found in other extensions of the Standard Model featuring similar ingredients, such as, e.g., the possibility of fermionic doublet-dominated dark matter. 

Finally, our results may be used as a guideline for future specific collider studies. It is to be noted that results from experimental searches for, e.g., supersymmetric partners of the Standard Model particles, may be used to obtain approximate exclusion limits on the present model. Such a study is, however, beyond the scope of the present paper and left for future work.

% ===========================================================================
\acknowledgments

The authors would like to thank W.~Porod for helpful discussions, in particular concerning the neutrino mass calculation and the {\tt SPheno} implementation of the model. B.\,H.\ thanks S.~Esch and M.~Klasen for valuable discussions in an early stage of this work. The work of M.\,S.\ is funded by a Ph.D.\ grant of the French Ministry for Education and Research. This work is supported by Campus France/DAAD, project PROCOPE 46704WF, and by {\it Investissements d’avenir}, Labex ENIGMASS, contrat ANR-11-LABX-0012. This research was supported by the Deutsche Forschungsgemeinschaft (DFG, German Research Foundation) under grant 396021762 - TRR 257. The plots presented in this article have been obtained using {\tt MatPlotLib} \cite{MatPlotLib}. 

% ===========================================================================
\appendix
% ======================================================
\section{Neutrino mass matrix}
\label{App:NeutrinoMassMatrix}
% ======================================================

As already stated in Sec.\ \ref{Sec:NeutrinoMasses}, the neutrino mass matrix can be decomposed into a part containing the couplings and a part containing the loop integrals, according to 
\begin{align}
    {\cal M}_{\nu} ~=~ {\cal G}^t \, M_L \, {\cal G} \,.
\end{align}
The matrix $M_L$ is related to the loop integrals according to Fig.\ \ref{Fig:RadiativeNeutrinoMasses}. In the present model, i.e.\ with two non-zero neutrino masses, this matrix has three independent components, which in the mass eigenbasis can be expressed as (see also Ref.\ \cite{Esch2018})
\begin{align}
    \begin{split}
    \left( M_L \right)_{11} ~&=~ \sum_{k,n} b_{kn}  \, (U_{\chi}^{\dagger})^2_{3k} \, (U_{\phi}^{\dagger})^2_{1n} \,, \\
    \left( M_L \right)_{12} ~=~ \left( M_L \right)_{21} ~&=~ \frac{1}{\sqrt{2}} \sum_{k,n} b_{kn} \, (U_{\chi}^{\dagger})_{1k} \, (U_{\chi}^{\dagger})_{3k} \, (U_{\phi}^{\dagger})_{1n} \, (U_{\phi}^{\dagger})_{2n}  \,, \\
    \left( M_L \right)_{22} ~&=~ \frac{1}{2} \sum_{k,n} b_{kn}  \, (U_{\chi}^{\dagger})^2_{1k} \, \Big[ (U_{\phi}^{\dagger})^2_{2n} - \big( U_{\phi}^{\dagger} \big)^2_{3n} \Big] \,,
    \end{split}
\end{align}
where the sums run over the neutral fermion mass eigenstates ($k=1,2,3$) and the neutral scalar mass eigenstates ($n=1,2,3$, the last one corresponding to the pseudo-scalar $A^0$). Moreover, the coefficients stemming from the loop integrals are given by
\begin{align}
    b_{kn} ~=~ \frac{1}{16\pi^2} \frac{m_{\chi^0_k}}{m_{\phi^0_n}^2 - m_{\chi^0_k}^2 } \biggr[ m_{\chi^0_k}^2  \ln\! \big( m_{\chi^0_k}^2 \big) - m_{\phi^0_n}^2 \ln\! \big( m_{\phi^0_n}^2 \big) \biggr]\,.
    \label{Eq:NeutrinoLoops}
\end{align}
Note that all divergences as well as the dependence on the renormalization scale vanish, as this is the leading-order contribution.

% ======================================================
\section{Casas-Ibarra parametrization}
\label{App:CasasIbarra}
% ======================================================

Here, we present the parametrization of the neutrino sector as introduced in Ref.\ \ref{Eq:CasasIbarra} applied to the case of the scotogenic model T1-2A under consideration in this work.

The matrix $M_L$ defined in Sec.\ \ref{App:NeutrinoMassMatrix} can be diagonalized as
\begin{align}
    M_L~=~ U_L^* \, D_L \, U^{\dagger}_L \,,
\end{align}
where $D_L$ contains the eigenvalues of $M_L$, and $U_L$ is the associated rotation matrix.

The neutrino mass matrix (see Eq.\ \eqref{Eq:NeutrinoMassMatrix}) is diagonalized by the PMNS matrix. The matrix $D_{\nu}$ containing the neutrino mass eigenvalues can then be expressed as
\begin{align}
    \begin{split}
    D_{\nu} ~&=~ {\rm diag}\big( m_{\nu_1}=0, m_{\nu_2}, m_{\nu_3} \big)
            ~=~ U_{\rm PMNS} \, {\cal M}_{\nu} \, U_{\rm PMNS}^t \\
            ~&=~ U_{\rm PMNS} \, {\cal G}^t \, M_L \, {\cal G} \, U_{\rm PMNS}^t 
            ~=~ U_{\rm PMNS} \, {\cal G}^t \, U_L \, D_L \, U^t_L \, {\cal G} \, U_{\rm PMNS}^t \,.
    \end{split}
\end{align}
The diagonal matrices $D_L$ and $D_{\nu}$ fulfill the relations $D_L = D^{1/2}_L D^{1/2}_L$, $D^{1/2}_{\nu} D^{1/2}_{\nu} = D_{\nu}$, and $D^{-1/2}_{\nu} D^{1/2}_{\nu} = {\rm diag}\big( 0, 1, 1 \big)$. Making use of these identities leads to
\begin{align}
    D^{-1/2}_{\nu} \, D_{\nu} \, D^{-1/2}_{\nu} ~=~ 
        D^{-1/2}_{\nu} \, U_{\rm PMNS} \, {\cal G}^t \, U_L \, D^{1/2}_L \, D^{1/2}_L \, U^t_L \, {\cal G} \, U^t_{\rm PMNS} \, D^{-1/2}_{\nu}
        ~\equiv~ R^t \, R \,,
\end{align}
with the definition $R ~\equiv~ D^{1/2}_L U^t_L {\cal G} U_{\rm PMNS} D^{-1/2}_{\nu}$. The fact that $R^t R = {\rm diag}\big( 0, 1, 1 \big)$ implies that $R$ can be written in terms of one single parameter,
\begin{align}
    R ~=~ \begin{pmatrix} 0 & \sqrt{1-r^2} & r \\ 0 & -r & \sqrt{1-r^2} \end{pmatrix} \,.
\end{align}

Putting everything together, the coupling matrix ${\cal G}$ is obtained as the product
\begin{align}
        {\cal G} ~=~ U_L \, D^{-1/2}_L \, R \, D_{\nu}^{1/2} \, U^{*}_{\rm PMNS} \,
\end{align}
where $D_{\nu}$ and $U_{\rm PMNS}$ are known from experiment, $U_L$ and $D_L$ depend only on parameters related to the scalar and fermion sectors, and the matrix $R$ is parametrized through $r \in [-1; 1]$.

% ===========================================================================
\bibliographystyle{JHEP}
\bibliography{paper}

% ===========================================================================
\end{document}